\documentclass[amsmath,amssymb,aps,prfluids]{revtex4}

\usepackage{graphicx}
\usepackage{subfig,enumerate}
\usepackage[mathscr]{euscript}
\usepackage{upgreek}
\usepackage{cancel}
\usepackage{dcolumn}
\usepackage{bm}
\usepackage{hyperref}
\usepackage{xcolor}

\begin{document}

\newcommand{\sm}[1]{{\color{black} #1}}
\newcommand{\smbis}[1]{{\color{black} #1}}
\newcommand{\remove}[1]{}
\newcommand{\Pe}{\mbox{Pe}}
\newcommand{\ub}{\mathbf{u}}
\newcommand{\eb}{\mathbf{e}}
\newcommand{\Ub}{\mathbf{U}}
\newcommand{\nb}{\mathbf{n}}
\newcommand{\dd}{\mathrm{d}}
\newcommand{\totd}[2]{\frac{\mathrm{d} #1}{\mathrm{d} #2}}
\newcommand{\pard}[2]{\frac{\partial #1}{\partial #2}}

\newcommand{\fp}[1]{\textcolor{blue}{#1}}

\title{Confined self-propulsion of \sm{an isotropic active colloid}}
\author{Francesco Picella}
\author{S\'ebastien Michelin}%
 \email{sebastien.michelin@ladhyx.polytechnique.fr}
\affiliation{%
 LadHyX, D\'epartement de M\'ecanique, CNRS – Ecole Polytechnique, Institut Polytechnique de Paris, 91128 Palaiseau Cedex, France\\
}%

\date{\today}

\begin{abstract}
To spontaneously break their intrinsic symmetry and self-propel at the micron scale, isotropic active \sm{colloidal particles and droplets} exploit the non-linear convective transport of chemical solutes emitted/consumed at their surface by the surface-driven fluid flows generated by these solutes. Significant progress \sm{was recently made to understand} the onset of self-propulsion and non-linear dynamics. Yet, most models ignore a fundamental experimental feature, namely the spatial confinement of the colloid, and its effect on propulsion. In this work, the self-propulsion of \sm{an isotropic colloid} inside a capillary tube is investigated numerically. A flexible computational framework is proposed based on a finite-volume approach on adaptative octree-grids and embedded boundary methods\sm{. This method} is able to account for complex geometric confinement, the nonlinear coupling of chemical transport and flow fields, and the precise resolution of the surface boundary conditions, that drive the system's dynamics. Somewhat counter-intuitively, spatial confinement promotes the colloid's spontaneous motion by reducing the minimum advection-to-diffusion ratio or P\'eclet number, $\Pe$, required to self-propel; furthermore, self-propulsion velocities are significantly modified  as the colloid-to-capillary size ratio $\kappa$ is increased, \sm{reaching a  maximum at fixed $\Pe$ for an optimal confinement $0<\kappa<1$}. These properties stem from a fundamental change in the dominant chemical transport mechanism with respect to the unbounded problem : with diffusion now restricted in most directions by the confining walls, the excess solute is predominantly convected away downstream from the colloid, enhancing front-back concentration contrasts. These results are confirmed quantitatively using conservation arguments and lubrication analysis of the tightly-confined limit, $\kappa\rightarrow 1$.

\end{abstract}

\maketitle

\section{Introduction}
Recent developments in the design of synthetic micro-swimmers open new opportunities for engineering and biomedical applications~\citep*{Nelson2010}. Popular designs closely follow locomotion strategies observed in Nature, such as beating flexible appendages~\citep{Dreyfus2005} or rotating chiral filaments~\citep{Magdanz2020}, breaking time-reversibility to ensure for the propulsion of such small-scales swimmers in viscous environments~\citep{Purcell1977,Lauga2009}.
But in contrast to their biological counterparts, these synthetic bio-mimetic swimmers essentially behave as \textit{marionettes}~\citep{Brooks2020}, relying on some external tether for both energy supply and motion control, such as \sm{magnetic, optic or acoustic fields~\citep{Koleoso2020,Bunea2019,Rao2015}}. Still, practical difficulties, such as miniaturisation and manufacturing of their moving parts, have so far hindered their use for practical applications.

Active colloids stem from a fundamentally-different paradigm, featuring no moving parts~\citep{Moran2019}. \sm{Just like bacteria or other swimming cells~\citep{Berg1993}, they are instead able to extract and convert} into motion,  energy  tapped directly from their immediate environment (e.g. non-uniform distribution of a physico-chemical properties) in a mechanism known as \textit{phoresis}~\citep{Anderson1989}. Beyond technological applications, active colloids have been central to the recent developments in the study of so-called \emph{active matter}, in an effort to understand and characterise the collective dynamics and self-organisation among large suspensions of microscopic self-propelled systems~\citep{Marchetti2013,Bechinger2016}.

Surface \emph{activity} of the colloid is the most popular approach to the generation of the local physico-chemical (e.g. solute) gradients required for propulsion, and can take the form of reactions catalysed by a surface coating~\citep{Howse2007}, encapsulated in a droplet~\citep{Thutupalli2011} or rely on micellar dissolution~\citep{Izri2014,Moerman2017}. Combined with a \emph{mobility}, namely the ability to convert local gradients along the surface into fluid motion or fluid stresses, this opens the way for the self-diffusiophoretic motion of chemically-active swimmers \sm{that are able to generate themselves the local gradients into which they subsequently propel~\citep{Golestanian2007,Moran2017,Maass2016}.}

\sm{The fundamental propulsion features are critically impacted by the transport of chemical solutes involved in phoresis within the fluid}, or more specifically by the ratio of convective transport and molecular diffusion, measured by the P\'eclet number, $\Pe$. Based on that measure, two different classes of active colloids can be distinguished. When $\Pe\ll 1$, solute transport is dominated by diffusion and is thus independent from the fluid (and colloid's) motion\sm{: this} is specifically the case of classic autophoretic particles, such as the canonical Au-Pt Janus colloids~\citep{Paxton2004}, that are typically micron scale and use small and rapidly-diffusing solutes~\citep[e.g. dissolved gases,][]{Moran2017}. In that case, generating gradients requires embedding some asymmetry in the design of the swimmer through inhomogenous surface activity~\citep{Paxton2004,Howse2007} or an anisotropic geometry~\citep{Kummel2013,Michelin2015}. \sm{This can also be achieved }through asymmetric assembly of isotropic colloids~\citep{Varma2018,Yu2018}.

In contrast, chemically-active droplets are relatively large (typically $10$--$100\mu$m in diameter) and their activity is based on their micellar dissolutions into the outer fluid phase~\citep{Maass2016,Morozov2020}. \sm{The} solutes exchanged at the droplet's surface and responsible for its propulsion are large molecular structures (surfactant, micelles...) and thus diffuse slowly in the fluid: advective effects are here non-negligible and ${\sm{\Pe=O(1)-O(100)}}$~\citep{Hokmabad2021}. Symmetry-breaking is achieved through an instability resulting from the non-linear convective transport of the solute species by the fluid flows  \sm{generated} from phoretic and Marangoni effects at the droplet surface~\citep{Izri2014,Morozov2019a}. In contrast with autophoretic particles with $\Pe\ll 1$, this non-linear hydro-chemical coupling provides the droplet with complex and tunable individual behaviour~\citep{Suga2018,Hokmabad2021}, and can even lead to the emergence of chaotic dynamics~\citep{Morozov2019b,Hu2019}.

The mechanism at the heart of the droplet's self-propulsion, i.e. the nonlinear feedback coupling between the flow and chemical fields, is mathematically and physically relevant regardless of whether the mobility stems from phoretic slip flows or Marangoni stresses, both emerging from tangential gradients in solute concentration~\citep{Michelin2013,Izri2014,Morozov2019b}. In fact, both mechanisms most likely co-exist in active droplets, whose surface is densely covered by surfactant species due to the saturation of the suspending fluid. \sm{Also, in experiments, active droplets remain spherical (the relevant capillary numbers are small) except when their radius is larger than the capillary or chamber size~\citep[see e.g.][]{deBlois2021}}. As a result, isotropic phoretic particles can be considered in a first approximation as the limit case of swimming droplets with large internal viscosity.

Despite their systematic presence in experimental settings, due to the droplets' non-neutral buoyancy~\citep{Kruger2016b,Cheon2021} \sm{or as a requirement for accurate quantitative measurements~\citep[e.g. confocal microscopy][]{Hokmabad2021}}, theoretical models most often ignore the presence of confining boundaries and focus on droplets in unbounded fluid domains, leaving unexplored their role on the emergence and persistence of self-propulsion. Recent experimental measurements have shown significant modifications of the flow field around the droplet when placed close to or between rigid walls~\citep{deBlois2019}, and theoretical modelling unveiled the non-trivial alterations of the hydro-chemical coupling induced by confinement~\citep{Lippera2020}. Beyond the influence of a single flat wall, recent experiments have also shown that self-sustained motion can also occur in strongly-confined settings, such as small capillary tubes~\citep{Illien2020,deBlois2021}.

Although few quantitative measurements or estimates can be found, active droplets are likely to evolve very close to their confining boundaries~\citep{Cheon2021}, in a regime where classical work on lubricating flows or  model micro-swimmers demonstrate that hydrodynamic drag~\citep{Kim1991} and self-propulsion velocities~\citep*{Zhu2013} are significantly modified in comparison with their characteristics in unbounded fluid domains. Significant changes in the self-propulsion of active droplets would therefore not be surprising.

The central goal of the present work is to provide a much needed insight on the sustained self-propulsion of such isotropic active particles or droplets in strongly-confined settings, i.e. inside a capillary tube. In the case of diffusion-dominated diffusiophoretic swimmers $(\Pe \rightarrow 0)$, the hydrodynamic and solute \sm{evolutions reduce} to sequential \emph{linear} Laplace and Stokes problems, for which a number of different numerical techniques are available, such as  Boundary Element Methods~\citep*{MontenegroJohnson2015} or two recent extensions of hydrodynamic solvers for the diffusive problem, based on  Stokesian dynamics~\citep{Yan2016} or the Force Coupling Method~\citep*{RojasPerez2021}. 

In contrast, the numerical simulation of instability-driven, isotropic autophoretic swimmers at non-zero $\Pe$ poses new and specific challenges due to the inherent nonlinearity of the problem \sm{in addition to} the presence of moving boundaries where chemical and hydrodynamic forcings are applied. Up to date, \sm{most} simulations considering the full non-linear hydrochemical coupling of active droplets rely on some truncated spectral expansion, mapped either onto cylindrical~\citep{Hu2019}, spherical~\citep{Michelin2013} or bi-spherical coordinates~\citep{Lippera2020,Lippera2020b}.  \sm{This approach is well-suited} for simple geometric configurations (e.g. unbounded flows, two-sphere interactions), but precludes the study of the dynamics of such swimmers placed under generic spatial confinement or even in a cylindrical pipe.

To overcome this hurdle, we present here a generic method to obtain the non-linear hydro-chemical dynamics of a single isotropic autophoretic particle under complex confinement using a novel approach based on \textit{embedded boundaries}~\citep{Johansen1998,Schwartz2006} and developed on top of the adaptive quadtree-octree flow solver \textit{Basilisk}~\citep{Popinet2015}. Our approach, based on a finite volume framework, does not require any \emph{a priori} assumption on the form of the hydrodynamic or chemical fields, nor on the number or shape of the solid boundaries, thus making it suitable for the study of  complex confinement geometries and/or collective particle/droplet dynamics.

The paper is organised as follows. Section~\ref{sec:description} introduces the physical problem considered, namely that of a single isotropic autophoretic particle swimming along the axis of a round capillary tube. The numerical technique used to solve the problem is then presented in Sec.~\ref{sec:numerics} together with several numerical validations.
The impact of spatial confinement, i.e. the relative radius of the capillary and particle, is then analysed in detail in Section~\ref{sec:results} using this numerical method. Using global conservation arguments and lubrication analysis, Sec.~\ref{sec:theory} then confirms theoretically the qualitative and quantitative evolution of the propulsion characteristics in the strong-confinement limit (i.e. tightly-fitting sphere).
Finally we summarize our findings and outline some perspectives on this work in Sec.~\ref{sec:conclusions}.

\section{Phoretic self-propulsion in a capillary}\label{sec:description}

We consider the dynamics of a single spherical phoretic particle of radius $a$, immersed in a Newtonian fluid of viscosity $\eta$ and density $\rho$, inside a circular capillary of radius $R$ and axis $\eb_z$. The particle is chemically-active and releases \smbis{or absorbs} a solute of \sm{concentration \smbis{$c^*$} and} molecular diffusivity $D$ into its fluid environment with a constant and isotropic flux $\mathcal{A}$ (activity), so that along the particle's boundary $\Gamma_p$
\begin{equation}\label{eq:activity}
\left.\sm{D}\nb\cdot\nabla \smbis{c^*}\right|_{\Gamma_p}=-\mathcal{A},
\end{equation}
with $\nb$ the unit outward normal. The short-ranged interaction of solute molecules with the \sm{particle surface} within a thin interaction layer of thickness $\lambda\ll a$ introduces an effective hydrodynamic slip \smbis{$\tilde\ub^*$} along the \sm{particle surface} in response to local tangential solute gradients~\citep{Anderson1989}
\begin{equation}\label{eq:phoretic_slip}
    \smbis{\mathbf{\tilde{u}}^*} = \mathcal{M} \nabla_s \smbis{c^*},
\end{equation}
with $\mathcal{M} \approx k_B T \lambda^2/\eta$ the phoretic \textit{mobility} of the particle, with $k_B T$ the thermal energy and $\nabla_s=(\mathbf{I}-\nb\nb)\cdot\nabla$  the tangential gradient operator projected onto the \sm{particle surface}. \sm{Note that taking \smbis{$\mathcal{M}$} as a constant characteristic property of the particle surface is valid for neutral solutes, but can also be valid when concentration contrasts are small enough~\citep{Anderson1989}.} 

The activity $\mathcal{A}$ and mobility $\mathcal{M}$ coefficients characterise the physico-chemical properties of the particle surface and can be positive or negative; from these, a characteristic phoretic velocity scale can be defined as $\mathcal{V}=|\mathcal{AM}|/D$.  Given the characteristic size and velocities of confined phoretic microswimmers~\citep{deBlois2019,Lippera2020,Hokmabad2021}, the \sm{fluid and colloid inertia} can be neglected, i.e. the \sm{Reynolds} number $\mbox{Re}=\rho\mathcal{V}a/\eta$ is negligible, so that the motion of the particle can be described using the steady Stokes equations.

In the following, all quantities of interest are made dimensionless  using $a, \mathcal{V}, a/\mathcal{V}$ and $a|\mathcal{A}|/\mathcal{D}$ as characteristic length, velocity, time and concentration, respectively.
The resulting dimensionless equations for the \sm{dimensionless flow velocity $\mathbf{u}$, pressure $p$ and concentration $c$}  are:
\begin{align}
    \nabla^2 \mathbf{u} = \nabla p, \quad \nabla \cdot \mathbf{u} = 0,\label{eq:stokes}\\
    \frac{\partial c}{\partial t} + \mathbf{u} \cdot \nabla c = \frac{1}{\Pe} \nabla^2 c,\label{eq:solute}
\end{align}
with $\Pe = |\mathcal{AM}|a/\mathcal{D}^2$, the P\'eclet number, which is a measure of the relative contribution of advection and diffusion to the transport of solute. The radius ratio, $\kappa=a/R\in[0, 1]$, is a measure of the confinement level and is the second key dimensionless parameter of the problem.

The relevant boundary conditions for the concentration field at the surface of the (active) particle $\Gamma_p$ and (inert) confining wall $\sm{\Gamma_d}$ are
\begin{equation}\label{eq:solute_BC}
   \left . \partial c / \partial n \right|_{\Gamma_p}= -A, \quad \left. \partial c / \partial n \right|_{\sm{\Gamma_d}} = 0,
\end{equation}
while, for the velocity field,
\begin{equation}\label{eq:velocity_BC}
    \left .\mathbf{u}\right|_{\Gamma_p} = \mathbf{\tilde{u}} + \mathbf{U} + \mathbf{\Omega} \times (\mathbf{x}-\mathbf{X}), \quad \left . \mathbf{u} \right|_{\sm{\Gamma_d}} = \mathbf{0}
\end{equation}
where $\Ub$ and $\boldsymbol\Omega$ are the translation and rotation velocities of the particle, $\tilde\ub=M\nabla_s c$ is the \sm{dimensionless} phoretic slip velocity at \sm{a general point $\mathbf{x}$ on} the particle surface and $\mathbf{X}$ is the position of the particle center. \sm{The phoretic mobility of the wall is neglected here in front of that of the active particle/droplet, but could easily be accounted for within the same framework~\citep[see e.g.][]{Sherwood2018}.} Here, \sm{ $A = \mathcal{A}/|\mathcal{A}| $ and $M=\mathcal{M}/|\mathcal{M}|$} denote the dimensionless \smbis{activity and mobility}. When $AM=-1$, no self-propulsion is observed for an isolated particle in unbounded flow~\citep{Michelin2013}; as our goal is to analyse the effect of confinement on self-propulsion, we consider  in the following that $A=M=1$. 

Finally, in the absence of any external force, the total hydrodynamic force and torque on the particle must vanish at all time,
\begin{equation}\label{eq:force_free}
    \mathbf{F} =\int_{\Gamma_p}\boldsymbol\sigma\cdot\nb\dd S=0,\qquad \mathbf{T} =\int_{\Gamma_p}\boldsymbol(\mathbf{x}-\mathbf{X})\times(\boldsymbol\sigma\cdot\nb)\dd S=0,
\end{equation}
with $\boldsymbol\sigma=-p\mathbf{I}+\nabla\ub+\nabla\ub^T$ the Newtonian stress tensor.

\section{Numerical solution}\label{sec:numerics}
\subsection{Axisymmetric problem and co-moving frame}

\begin{figure}
    \centering
    \includegraphics[width=0.8\textwidth]{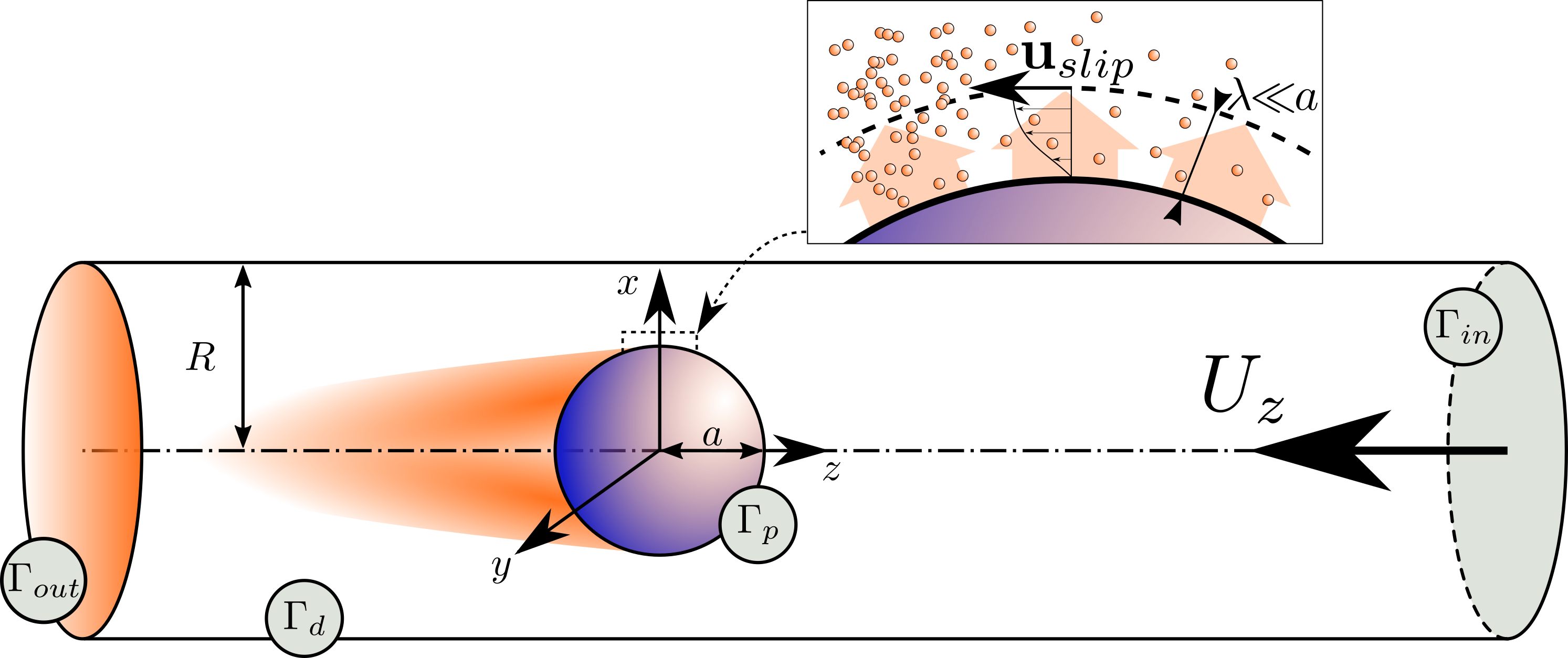}
    \caption{Self-propulsion of a single isotropic phoretic particle of radius $a$ along the axis of a cylindrical pipe of radius $R$ (viewed here in the reference frame of the particle). The particle-to-pipe radius ratio, $\kappa=a/R$, is a measure of the level of confinement. The particle and pipe surfaces are noted $\Gamma_p$ and $\Gamma_d$, respectively. $\Gamma_\textrm{in}$ and $\Gamma_\textrm{out}$ denote cross sections of the pipe far ahead and behind the particle, respectively.
    (Inset) Within a thin interaction layer of thickness $\lambda$, local surface gradients in the chemical solute (orange) released from the particle surface induce a net hydrodynamic slip. 
    }
    \label{fig:sketch}
\end{figure}

In the following, we will focus on the axial self-propulsion of the particle, for which the problem remains completely axisymmetric. In steady state, the concentration and velocity fields are time-independent when measured in a reference frame moving with the particle. For convenience (e.g. to avoid any need for re-meshing of the computational domain), we analyse the problem in that co-moving reference frame, where the particle is fixed, and the boundary conditions for the velocity field become
\begin{equation}\label{eq:stokes_BC} 
    \left. \mathbf{u}\right|_{\Gamma_p} = \mathbf{\tilde{u}}, \quad \left. \mathbf{u}\right|_{\sm{\Gamma_d}} = -U_z \smbis{\mathbf{e}_z},
\end{equation}
where $U_z$ is the physical velocity of the particle relative to the wall in the laboratory frame, and completely determines the steady self-propulsion dynamics of the particle along the confining tube's axis \sm{(see Fig.~\ref{fig:sketch})}. 

It should be noted nevertheless that the numerical methodology presented in the following can be generalised to non-axisymmetric and unsteady configurations. Unsteady simulations of the particle's dynamics in the laboratory frame \sm{were also performed with non-axisymmetric initial conditions (i.e. particle position, direction and intensity of the particle velocity) and showed }that this axisymmetric self-propulsion state is a stable attractor for the problem when $\Pe<15$, \sm{for all $\kappa$}, i.e. when the particle is released initially away from the axis, it relaxes after a transient to either a stationary state \sm{on the axis} or steady propulsion along the axis), establishing the physical relevance of the axisymmetric setting considered here.

\subsection{Numerical Method}\label{eq:numerical}
Equations \eqref{eq:stokes}, \eqref{eq:solute} and \eqref{eq:force_free}, with boundary conditions, Eqs.~\eqref{eq:solute_BC} and \eqref{eq:stokes_BC}, form a fully-coupled set of nonlinear PDE's problem. We solve these equations numerically in a cylindrical domain of length $L\gg R$ with the particle located at its center. 

Boundary conditions must be prescribed for the solute and flow fields on the upstream and downstream cross-sections of the computational domain, $\Gamma_\textrm{in}$ and $\Gamma_\textrm{out}$ \sm{located respectively at $z=\pm L/2$ from the center of the particle}. In the lab frame, the fluid is expected to be at rest with a homogeneous concentration of solute, far enough upstream and downstream of the particle, so that, \sm{in the reference frame co-moving with the particle},
\begin{equation}
\label{eq:BC_inout}
\left.\ub\right|_{\Gamma_\textrm{in},\Gamma_\textrm{out}}=-U_z\mathbf{e}_z, \left.\qquad \pard{c}{z}\right|_{\Gamma_\textrm{in},\Gamma_\textrm{out}}=0,
\end{equation}
\sm{with $U_z$ the \emph{a priori} unknown particle velocity, which is determined as part of the solution by enforcing the force-free condition on the particle.}

 As for the inertial fluid-solid coupling in high-Reynolds configurations~\citep{Selcuk2020}, the presence of the nonlinear advective coupling in the solute transport equations prevents the use of other popular numerical techniques such as multipole expansion~\citep{Sangani1996}, Boundary Elements Methods~\citep{Pozrikidis1992,MontenegroJohnson2015} or the Force Coupling Method~\citep{Delmotte2015,RojasPerez2021}, which are particularly suitable for purely diffusive problems. 
In such a limit, a detailed knowledge of the flow and concentration fields in the domain bulk (i.e. away from the computational domain boundary $\Gamma=\Gamma_p\cup\sm{\Gamma_d}\cup\Gamma_\textrm{in}\cup\Gamma_\textrm{out}$) is unnecessary to obtain the particle dynamics. In contrast, when $\Pe\neq 0$, the presence of the advective contribution to the solute transport, $\mathbf{u} \cdot \nabla c$, which is key to the \sm{understanding} and capture of the spontaneous self-propulsion of isotropic phoretic particles and droplets~\citep{Michelin2013,Izri2014}, imposes a change in the resolution paradigm, by requiring to determine $\ub$ and $c$ everywhere in the computational domain, and an accurate numerical treatment of this non-linear term in the solute transport equation.

We present here a novel approach to solve for the diffusiophoretic propulsion based on Basilisk, a popular open source framework for computational fluid dynamics~\citep{Popinet2015}. Borrowing techniques developed for high-$\mbox{Re}$ flow simulations, the non-linear diffusiophoretic problem is split into multiple sub-problems. The equations of evolution for the solute and flow fields are solved using finite volumes on hierarchically-arranged, adaptive quadtree/octree grids~\citep{Popinet2003}. To adapt to the Basilisk framework most efficiently, the hydrodynamic problem is described by the unsteady Stokes equations with a small Reynolds number ($\mbox{Re}=0.05$). To reach the steady state solutions, the hydrodynamic solver is called iteratively on a pseudo-time $\tilde{t}$, until the residuals between two pseudo-timesteps reaches a convenient threshold, i.e. ${\left| \mathbf{u}(\tilde{t}+\Delta\tilde{t}) - \mathbf{u}(\tilde{t}) \right| \lesssim 10^{-6} \left| \mathbf{u}(\tilde{t}) \right| }$, which generally takes $\mathcal{O}(10)$ successive calls. Note that this step represents the most time-consuming part of the method.
Stokes equations are solved using an operator-splitting method~\citep*{Bell1989}, with a viscous step (Poisson solver) followed by a projection onto a divergence-free space (Helmholtz solver). For the solute transport, Eq.~\eqref{eq:solute}, the diffusive Laplacian term is  handled implicitly while the advective contribution is computed using the Bell-Colella-Glaz (BCG) second-order upwind method~\citep{Bell1989}.

The description of all solid-fluid interfaces that do not match a rectangular mesh is based on the method of embedded boundaries~\citep{Johansen1998,Schwartz2006}, allowing for a second-order accurate computation of the additional fluxes to be included in the finite-volume balance in order to enforce a prescribed boundary condition within cells containing a fluid-solid interface $\Gamma_p$ or $\sm{\Gamma_d}$~\citep{Schneiders2016}. 
 Hydrodynamic forces are then computed \textit{a posteriori} by numerical integration of the stress tensor on the \sm{particle surface}.

At this point, we dispose of an efficient numerical framework for the computation of the flow velocity and solute concentration, $(\mathbf{u},p, c)$, for boundaries of any shape, for a given particle velocity. The dynamics of the particle (here completely characterised by $U_z$, its axial velocity) is further determined through the instantaneous force-free constraint, which writes here simply as $F_z=0$.
The Stokes problem is linear regardless of the confinement level $\kappa$; therefore the axial force on the particle is an affine function of the solid body translation $U_z$, for a given slip velocity $\tilde\ub$, i.e.
\begin{equation}\label{eq:affine_main}
    F_z(\mathbf{\tilde{u}},U_z) = \mathcal{R} U_z + \mathcal{Q}(\tilde\ub),
\end{equation}
with $\mathcal{R}$ the axial drag coefficient on a rigid sphere translating along the axis of the cylindrical pipe, and $\mathcal{Q}$ a scalar that is completely determined by the surface slip and the level of confinement. Both $\mathcal{Q}$ and $\mathcal{R}$ are independent of $U_z$ and solely depend on geometry \sm{(and on the slip velocity in the case of $\mathcal{Q}$)}, and are determined numerically as follows. At each time step, the Stokes problem is solved twice \sm{for the same slip velocity $\tilde{\mathbf{u}}$}: (i) for the real problem, using a first guess of the swimming speed $U_z$ from the previous time step, and (ii) for an auxiliary problem with a different \sm{and arbitrary} $U_z^\textrm{aux}$. For each, the corresponding axial force on the particle is computed and, using both solutions together with Eq.~\eqref{eq:affine_main} provides $(\mathcal{Q},\mathcal{R})$ from which the correct swimming speed satisfying the force-free condition is obtained as $U_z(F_z=0)=-\mathcal{Q}/\mathcal{R}$. 

A cubic domain of size $L/a=128$ was used on an adaptive mesh refinement, with the finest spatial discretization reaching $32$ computational cells per particle radius $a$, with $\approx 2000$ cells describing the \sm{particle surface}. The fluid domain (Figure~\ref{fig:sketch}) is then cut out from \sm{this} cubic volume employing the embedded boundaries approach~\citep{Bell1989} and the particle is set in the origin of the coordinate system, placed in the centre of the computational cuboid. \sm{Mesh is automatically adapted so as to ensure that maximum spatial refinement is always ensured on top of solid-liquid interfaces. Elsewhere, mesh is refined (resp. coarsened) whenever velocity or solute gradients is more (resp. less) than a prescribed threshold using an adaptive wavelet algorithm~\citep[see e.g.][]{vanHooft2018}.} \smbis{Using this approach and comparing the results for maximum spatial discretisations of 32 and 64 cells per unit length, we obtained a match in both swimming velocity and solute concentration fields, with a typical discrepancy on the swimming velocity lower than 0.1\% ($\mbox{Pe}=6$, $\kappa=0.5$) and reaching a maximum $2\%$ discrepancy for the most confined case considered ($\Pe=6$, $\kappa=0.9$).}

\subsection{Validation}\label{sec:validation}

We now proceed with the validation of the proposed framework and algorithms testing the main physical features against classical literature cases, namely (i) the self-propulsion of a model micro-organism using a prescribed surface slip (i.e. a so-called squirmer) in strong spatial confinement~\citep{Zhu2013} and (ii) the self-propulsion of isotropic particles due to non-linear hydro-chemical coupling~\citep{Michelin2013}. The first case, for which the hydrodynamic slip is imposed, allows for the validation of the hydrodynamic solver and the enforcement of the force-free constraint, while the second provides a validation of the coupled hydro-chemical solver.

\subsubsection{Squirmer in a pipe}
The behaviour of a single squirmer inside a cylindrical pipe for different level  of confinement is considered, as studied in Ref.~\cite{Zhu2013} using a boundary element method. A steady slip velocity $\tilde\ub$ is imposed on the particle surface, which corresponds to a neutral squirmer, which would swim at a velocity $U_z^*\mathbf{e}_z$ in the absence of any confinement, namely:
    \begin{equation}\label{eq:squirmer}
    \tilde{\mathbf{u}}^\textrm{squirmer}=  -\dfrac{2U_z^{*}}{3} (\mathbf{I}-\nb\nb)\cdot\mathbf{e}_z.
    \end{equation}
For an unconfined case, the algorithm recovers  within $\pm 0.1\%$ the swimming velocity predicted by the reciprocal theorem~\citep{Stone1996}, i.e. the average of the surface slip velocity on the particle surface. For confined cases, with $0<\kappa\leq 0.5$, the maximum relative error between the present results and that of Ref.~\cite{Zhu2013} is less than $2\%$ (Table~\ref{tab:confined_squirmer_comparison}).
\begin{table}
\centering
\begin{tabular}{cccc}
$\kappa=a/R$ & Zhu \textit{et al.}~\citep{Zhu2013} & present work & relative error ($\%$) \\\cline{1-4}
0.2 & 0.984 & 0.983 & 0.102 \\
0.3 & 0.948 & 0.943 & 0.530 \\
0.4 & 0.884 & 0.872 & 1.376 \\
0.5 & 0.791 & 0.776 & 1.933 \\
\end{tabular}
\caption{\label{tab:confined_squirmer_comparison} Steady-state swimming speed of a squirmer along the axis of a capillary tube for varying confinement ratio $\kappa$ and comparison with the results of Ref.~\cite{Zhu2013}. The  swimming velocity is normalised by that in unbounded fluid domains.}
\end{table}
Physically, for the axisymmetric configurations tested here, the squirmer's swimming velocity is observed to decrease with increasing confinement. 
    
\subsubsection{Autophoretic propulsion in an unbounded domain}

The second comparison allows for the validation of the hydro-chemical solver in the absence of confinement ($\kappa\ll 1$), and in particular of the treatment of the nonlinear advective coupling of the Stokes and chemical problems, which is the essential ingredient of the spontaneous autophoretic motion studied here. The results are then compared to those of Ref.~\cite{Michelin2013} for strictly unbounded domains.
Steady self-propulsion is observed beyond a critical $\Pe$ after a transient, with a constant non-zero swimming speed as depicted in figure~\ref{fig:transient}. A detailed comparison with the results of Ref.~\citep{Anderson1982}{Michelin2013} shows that the present method is able to recover the correct swimming velocity  with an error lower than ${\approx 1\%}$ for the resolution considered (Table~\ref{tab:michelin_2013c_validation}).

\begin{table}
\centering
\begin{tabular}{cccc}
$\Pe$ & Michelin \textit{et al.}~\citep{Michelin2013} & present work & relative error ($\%$)\\\cline{1-4}
 4.0 & 0.0     & 0.0     & 0.00 \\
 5.0 & 0.04672 & 0.04670 & 0.043 \\
 6.0 & 0.06652 & 0.06649 & 0.045 \\
 7.5 & 0.08342 & 0.08387 & 0.53 \\
10.  & 0.08671 & 0.08782 & 1.26 \\
12.5 & 0.08333 & 0.08360 & 0.33 \\
15.0 & 0.07902 & 0.07953 & 0.64 \\
\end{tabular}
\caption{
    \label{tab:michelin_2013c_validation}
Steady-state swimming velocities as a function of  $\Pe$, for an isotropic autophoretic particle in an unbounded domain, and comparison with the results of Ref.~\cite{Michelin2013} for infinite domains.
}
\end{table}

\section{Axisymmetric self-propulsion inside a capillary}\label{sec:results}
\subsection{Unsteady vs. steady-state self-propulsion}

In practice, the coupled equations for the solute concentration and fluid velocity are integrated numerically in time  for fixed values of the P\'eclet number $\Pe$ and confinement ratio $\kappa$. For $0\leq \Pe\leq 15$, at long time, the particle propels at a steady velocity along the capillary axis: an axisymmetric steady-state is therefore reached in the reference frame of the particle for the solute concentration and flow fields, and we specifically focus here on the characterisation of such axisymmetric steady states.

\begin{figure}
    \centering
    \includegraphics[width=0.7\textwidth]{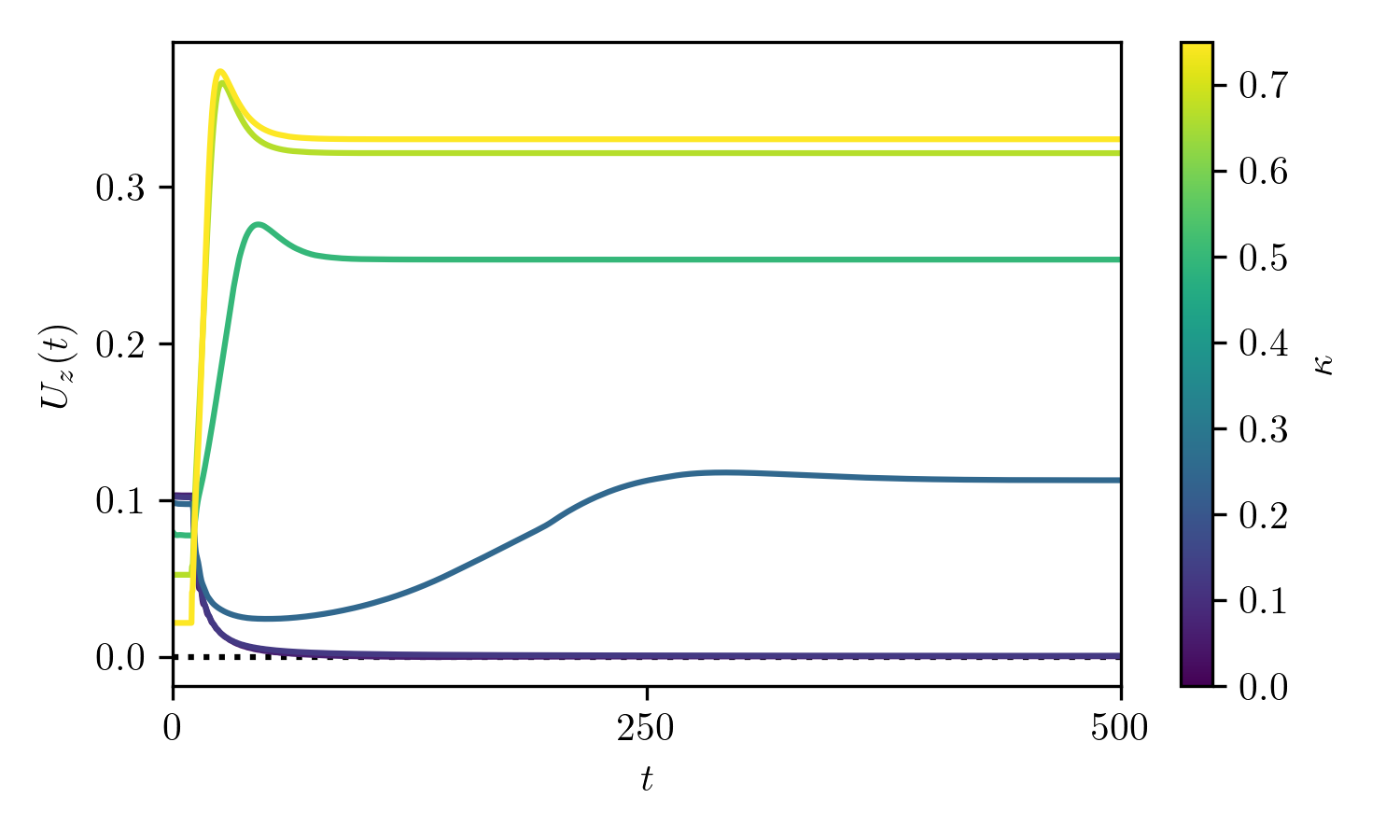}
    \caption{Unsteady axisymmetric propulsion velocity $U_z(t)$ of an isotropic phoretic particle along the capillary's axis for increasing level of confinement $\kappa$ (color) and $\Pe=2.5$. \sm{The results are reported for $\kappa=1/16$, $1/8$, $1/4$, $1/2$, $2/3$ and $3/4$.} During the initialisation phase ($t<t_{s}$ \sm{with $t_s=25$}), the slip velocity on the particle surface is imposed  (neutral squirmer), Eq.~\eqref{eq:squirmer}; for  $t>t_{s}$, the phoretic slip velocity  and particle dynamics are computed based on the actual surface concentration distribution of the solute, Eq.~\eqref{eq:phoretic_slip}, and the force-free condition on the particle.}
    \label{fig:transient}
\end{figure}

The simulation is initialised by prescribing during a short initialisation phase ($0\leq t\leq t_s$ with $t_s=25$ in non-dimensional units)  a fixed axisymmetric slip velocity on the surface of the particle, corresponding to a neutral squirmer with intrinsic swimming velocity $U_z^*=0.1$. For $t\geq t_s$, the true phoretic slip is computed directly from the actual concentration distribution and imposed at the particle surface. Such a procedure allows us to perturb the system and break the left-right symmetry. 
The resulting evolution of the swimming velocity is shown on Figure~\ref{fig:transient} for $\Pe=2.5$ and for increasing level of confinement. The self-propulsion of the particle during the initialisation (squirmer) phase decreases with $\kappa$, in agreement with the results of Ref.~\cite{Zhu2013}; \sm{indeed, in Figure~\ref{fig:transient}, the velocity $U_z$ in the initialisation phase ($t\leq 25$) where the slip velocity is imposed, is observed to be lower for larger values of $\kappa$ (tighter confinement).} 

For $t\geq t_{s}$, once the actual phoretic slip condition is enforced, the particle  relaxes after a transient toward its steady-state dynamics. Unless indicated otherwise, we thus refer to $U_z$ as the steady-state self-propulsion velocity of the particle when $t\gg t_s$. Two fundamentally different types of steady-state dynamics are observed on Figure~\ref{fig:transient} for $\Pe=2.5$, depending on the level of lateral confinement $\kappa$ of the particle. For weak confinement (i.e. small $\kappa$), the particle slows down and eventually comes to a stop; this is consistent with $\Pe=2.5$ being lower than the critical threshold for self-propulsion in unbounded domains ($\Pe_c=4$ for $\kappa=0$)~\citep{Michelin2013}. Note, that a steady state is reached for the particle velocity, flow field and concentration gradients, but not the average concentration which keeps increasing in time due to the fixed emission of solute at the particle surface and the confinement of the particle by chemically-inert walls. In contrast, for $\kappa\geq 0.2$, the particle maintains a net velocity that increases with $\kappa$ and saturates for the strongest confinements considered ($\kappa\approx 0.8$). \sm{Note that changing the magnitude of the initialisation velocity or the duration of the initialisation phase only modified the transient regime past $t\geq t_s$, but did not alter the nature of the observed steady state (i.e. fixed or self-propelled particle).}

\begin{figure}
    \centering
    \includegraphics[width=1.0\textwidth]{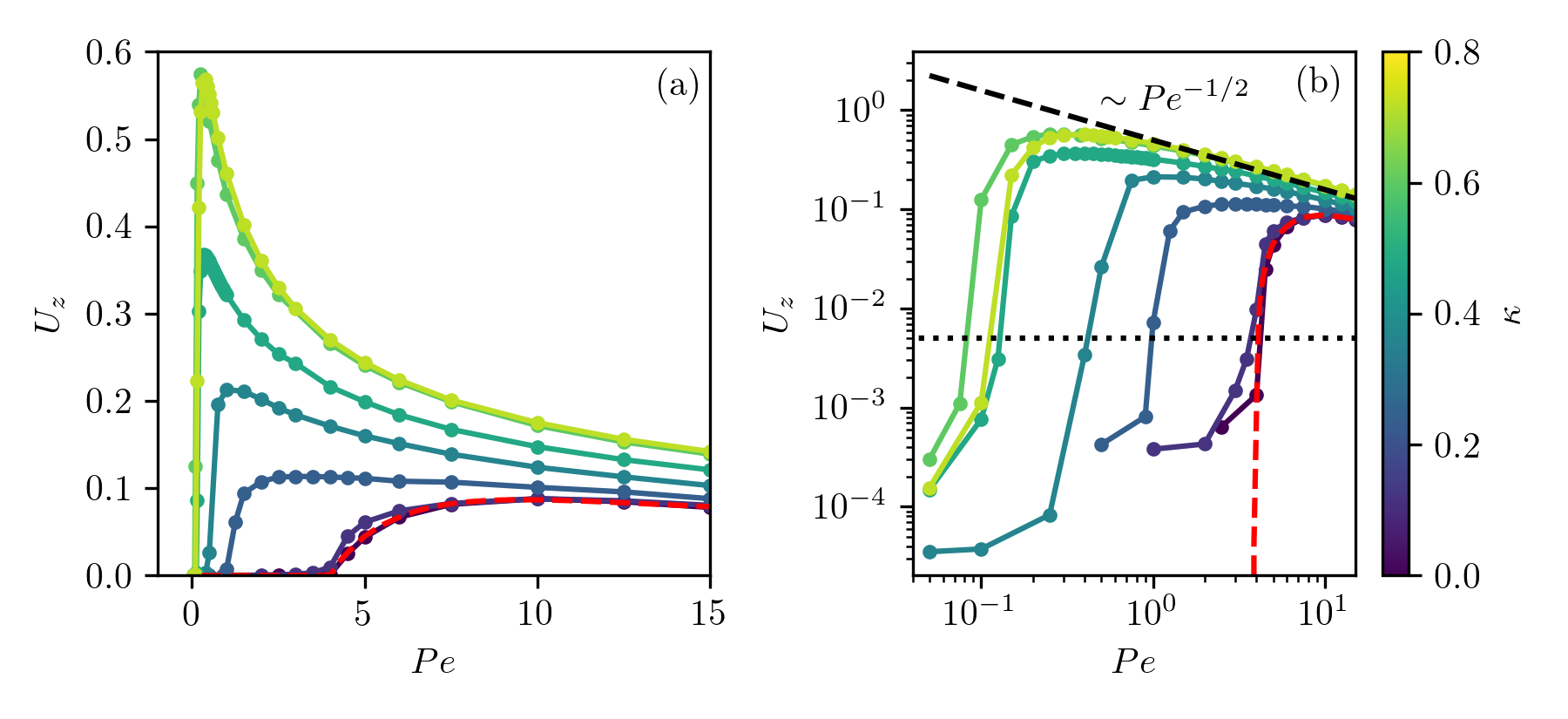}
    \caption{(a) Steady state axisymmetric swimming velocity $U_z$ of an isotropic active particle as a function of the convection-to-diffusion ratio, $\Pe$, and for increasing confinement $\kappa$ (color). (b) Same as (a) in logarithmic scale. The results for a single isolated particle in unbounded domains~\citep[$\kappa=0$,][]{Michelin2013} is shown for comparison (red dashed line). In (b), the dotted black line indicates the minimum velocity to determine the emergence of a net propulsion. \sm{The results are reported here for $\kappa=1/16$, $1/8$, $1/4$, $3/8$, $1/2$, $2/3$ and $3/4$.}
    }
    \label{fig:Pipe_U_Pe_kappa}
\end{figure}

\subsection{Self-propulsion velocity and critical threshold}
\label{sec:velocity}
In the following, we focus on the evolution of this steady-state self-propulsion and the influence of the proximity of the confining walls. To that end,  for ${0\leq\kappa\leq 0.8}$ and ${0<\Pe\leq 15}$, we systematically run time-dependent simulations until  a steady state is reached, with a constant swimming speed along the axis of the capillary. The results for $U_z(\Pe,\kappa)$ are reported on Figure~\ref{fig:Pipe_U_Pe_kappa}, and demonstrate the strong influence of confinement and an increase  of the self-propulsion velocity with confinement $\kappa$ for all $\Pe$. This effect is significant provided the distance to the wall is of the order of a few particle radii ($\kappa\gtrsim 0.2$),  confirming experimental observations~\citep{deBlois2021}. 

Beyond a systematic increase of the swimming velocity, Figure~\ref{fig:Pipe_U_Pe_kappa} also demonstrates several  other important features. Most importantly, confinement effects are strongest for low-to-moderate values of $\Pe$. We first note a significant reduction with $\kappa$ of the critical self-propulsion threshold $\Pe_c$. 
Furthermore, the presence of confinement strongly affects the evolution of $U_z(\Pe)$: in weakly-confined configurations, the velocity \sm{varies non-monotonically with $\mbox{Pe}$, and} increases smoothly from the threshold until it saturates for $\Pe\approx 10$ -- $20$ \sm{and decreases as $\mbox{Pe}$ is increased further~\citep{Michelin2013}}. In contrast, the velocity of strongly-confined particles ($\kappa\gtrsim 0.5$) scales as $1/\sqrt{\Pe}$ for most of the parameter range except  in the immediate vicinity of the threshold $\Pe_c(\kappa)$ where it increases sharply with $\Pe$ (Fig.~\ref{fig:Pipe_U_Pe_kappa}b).                                         As a result, the maximum swimming velocities are observed at low P\'eclet in strongly-confined environments (Figure~\ref{fig:Pipe_U_Pe_kappa}a).
Note that self-sustained motion is never observed for $\Pe=0$, regardless of $\kappa$: as for unbounded environments, convective transport of the solute by the phoretic flows is essential to the propulsion of isotropic particles, as it provides the required symmetry-breaking mechanism~\citep{Michelin2013,Izri2014}.

\begin{figure}
    \centering
    \includegraphics[width=1.0\textwidth]{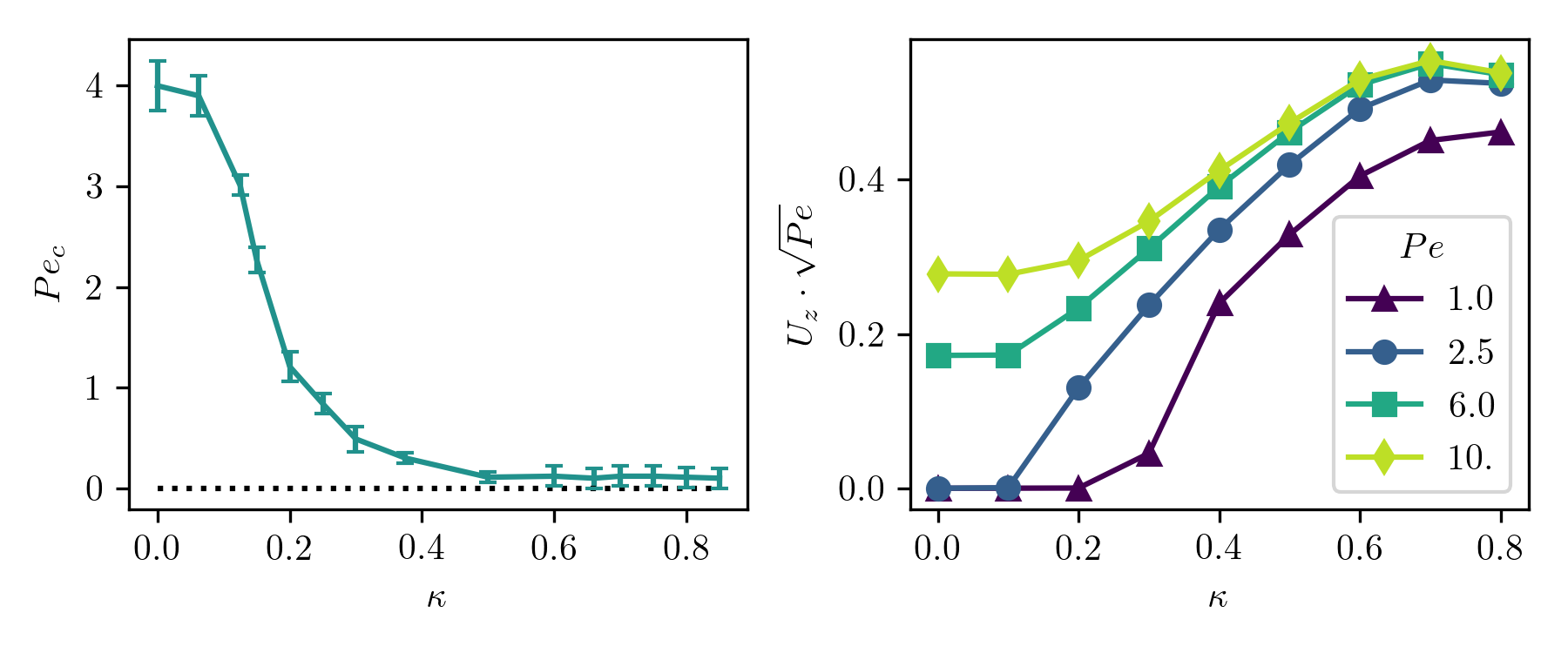}
    \caption{(a) Evolution with confinement ($\kappa$) of the critical threshold $\Pe_c$ for the onset of propulsion. For fixed $\kappa$, the numerical uncertainties on $\Pe_c$ are defined using the smallest (resp. largest) value of $\Pe$ for which the steady state velocity is greater (resp. smaller) than a numerical threshold $U_z^*=5\cdot 10^{-3}$. (b) Evolution of the rescaled particle velocity with confinement $\kappa$.}
    \label{fig:Pe_critic_kappa}
\end{figure}

Stronger confinement significantly promotes self-propulsion, by reducing the minimum P\'eclet number, $\Pe_c$ required for self-sustained autophoretic motion: while $\Pe_{c}=4$ for $\kappa=0$, the existence of a minimum $\Pe$ for self-propulsion persists throughout the range of confinement investigated but this threshold drops quickly as $\kappa$ is increased, with $\Pe_c\approx 0.1$ for $\kappa\gtrsim 0.5$ (Figure~\ref{fig:Pe_critic_kappa}). However, with the present numerical approach, it is not possible  to analyse the lubrication limit with significant precision to conclude on the asymptotic behaviour of $\Pe_c$ when $1-\kappa\ll 1$, and this asymptotic limit would require further analysis using a different approach.%

Except for rare examples~\citep{Hokmabad2021}, the Péclet number is fixed  in most experimental systems, and only spatial confinement can be controlled. For this reason, we also report on figure \ref{fig:Pe_critic_kappa}(b) the evolution of the rescaled swimming velocity for fixed $\Pe$ and variable confinement $\kappa$. In all cases, this rescaled representation (where we account for the dominant $\Pe^{-1/2}$ scaling of the velocity, see also Section~\ref{sec:theory}) demonstrates a non-monotonic evolution of the swimming velocity with $\kappa$, with a maximum at $\kappa\approx 2/3$, before entering the lubrication regime. This behaviour and its origin will be further discussed in Section~\ref{sec:scaling}.

\subsection{Effect of confinement on the solute distribution}\label{sec:concentration}
\begin{figure}
    \centering
    \includegraphics[width=1.0\textwidth]{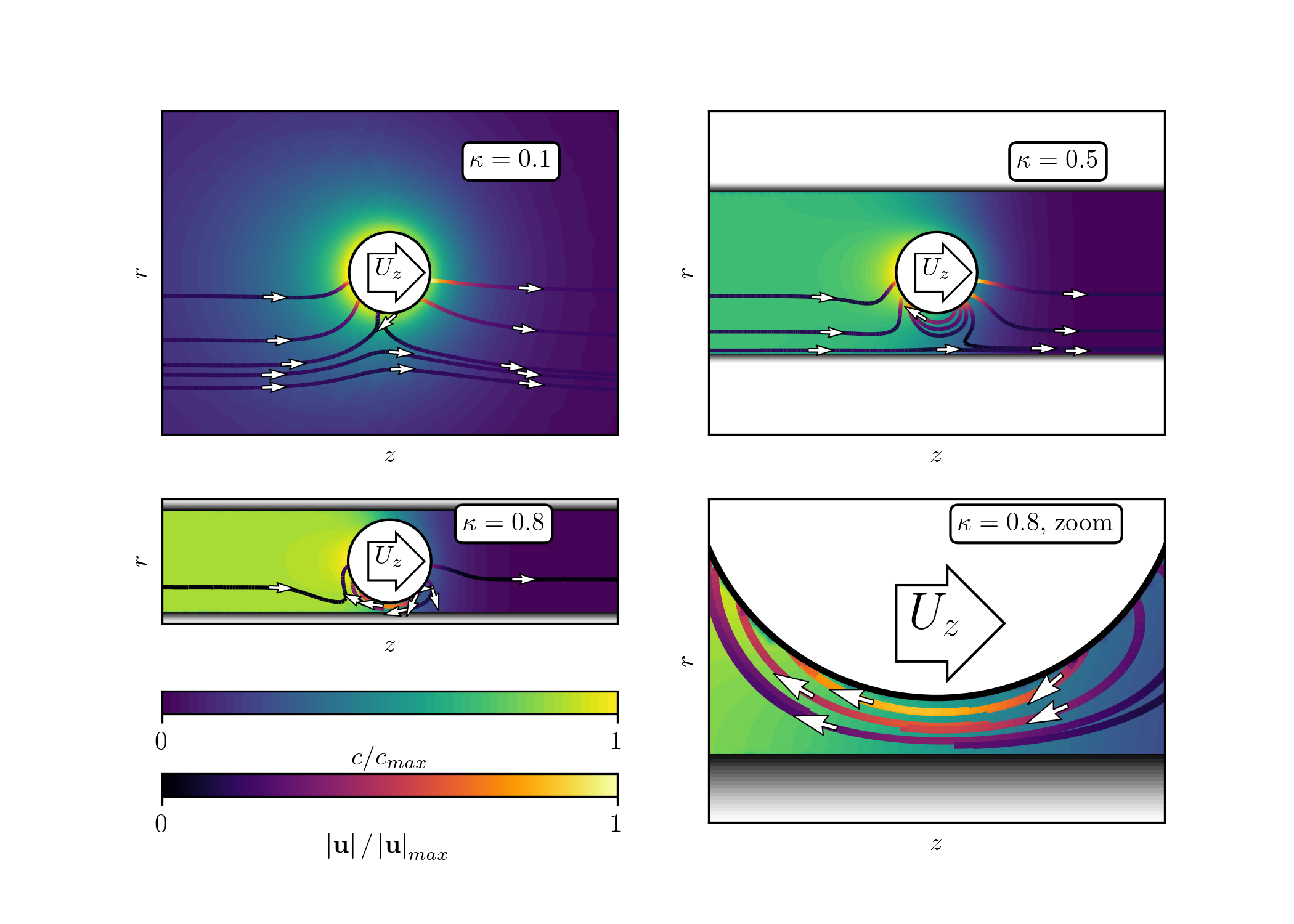}
    \caption{Steady state relative solute concentration distribution and representative streamlines (color-coded by the fluid velocity magnitude) around an isotropic phoretic particle in axisymmetric confinement for $\Pe=6$ and increasing $\kappa$ (in the \sm{laboratory} reference frame\sm{, i.e. fixed with respect to the capillary walls}).}
    \label{fig:pipe_concentration_streamplot}
\end{figure}
The peculiar evolution of the swimming velocity with confinement and its enhancement at low $\Pe$ is analysed by considering the detailed variations of the solute concentration around the isotropic phoretic particle in confined steady-state regimes (Figure~\ref{fig:pipe_concentration_streamplot}). 
For fixed $\Pe$, the solute distribution around the particle is fundamentally modified by confinement. 

In unbounded domains and for weak confinements, the solute distribution is characterised by a monotonic decrease in all radial directions around the particle, with a small front-back asymmetry maintained by the self-generated phoretic flows (Figure~\ref{fig:pipe_concentration_streamplot}, $\kappa=0.1$): in that case, the solute production at the particle surface is predominantly balanced by its radial diffusion away from the particle.

In contrast, for stronger levels of confinement (e.g. Figure~\ref{fig:pipe_concentration_streamplot}, $\kappa=0.8$),   lateral diffusion of the solute away from the particle is prevented by the lateral inert wall $\sm{\Gamma_d}$: in that case, the solute production by the particle's catalytic surface is predominantly balanced by its downstream convective transport by the phoretic flows. As a result, the capillary upstream from the particle is essentially solute-free, and the solute concentration saturates downstream from the particle at a much larger and uniform value. The largest solute concentrations are therefore found downstream and away from the particle, rather than on its surface as for the unbounded configuration.

A more detailed observation for strong confinement reveals that far upstream and downstream from the particle, the solute concentration becomes homogeneous due to the rapid lateral diffusion across the capillary (Figure~\ref{fig:pipe_concentration_streamplot}, $\kappa=0.8$). Additionally, as confinement is increased, the fluid layer separating the particle from the wall becomes very thin and chemical diffusion across this thin gap becomes dominant over other solute transport mechanisms: as a result, the solute concentration is homogenised across the whole fluid layer, despite the steady emission of solute from the particle surface (Figure~\ref{fig:pipe_concentration_streamplot}, $\kappa=0.8$, zoom).

\begin{figure}
    \centering
    \includegraphics[width=1.0\textwidth]{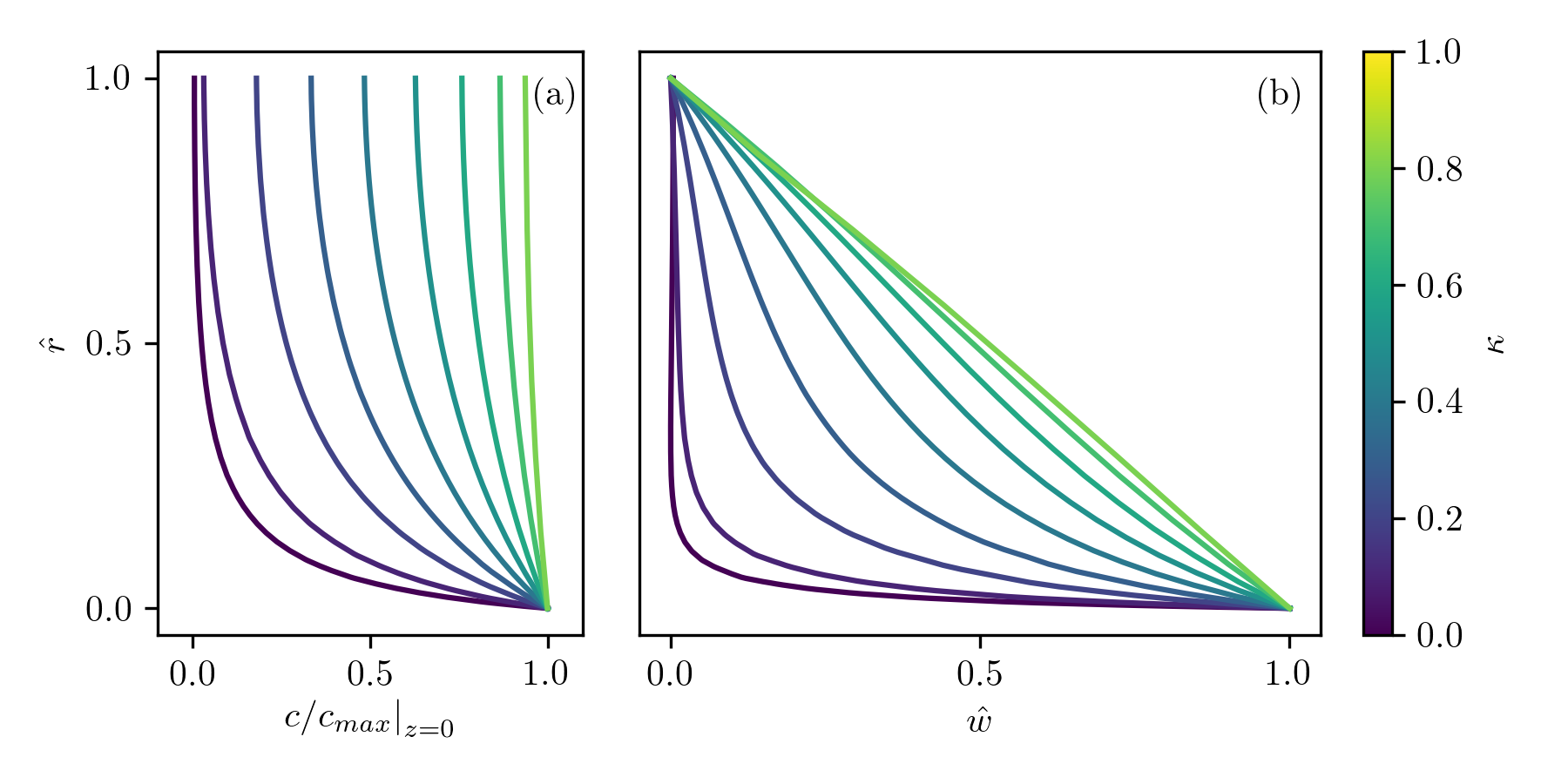}
    \caption{Relative spanwise distribution of (a) solute concentration and (b) streamwise flow velocity (in the particle's reference frame) within the fluid gap between the swimmer and the fixed walls ($z=0$) for increasing confinement $\kappa$ (color) and $\Pe = 6.0$. The rescaled radial variable $\hat{r}=(r-a)/(R-a)$ is defined such that $\hat{r}=0$ (resp. $\hat{r}=1$) corresponds to the particle (resp. wall) surface for all values of $\kappa$. In (b), the rescaled fluid velocity with respect to the particle is $\hat{w}=(w+U_z)/(w_\textrm{slip}+U_z)$ (note: here $w<0$ throughout the gap in the particle reference frame, and $w=-U_z$ at the wall, $\hat{r}=1$, while $w=w_\textrm{slip}<0$ at the particle surface, $\hat{r}=0$). \sm{The results are reported for $\kappa=n/10$ with $1\leq n\leq 9$.} 
    }
    \label{fig:spanwise_gap}
\end{figure}

This last observation is further confirmed quantitatively by the detailed evolution of the distribution of the concentration across the thin fluid layer (Figure~\ref{fig:spanwise_gap}a). While the solute concentration is only significant near the surface of the particle for small $\kappa$, the distribution of solute across the gap is uniform when $\kappa\rightarrow 1$.

\begin{figure}
    \centering
    \includegraphics[width=1.0\textwidth]{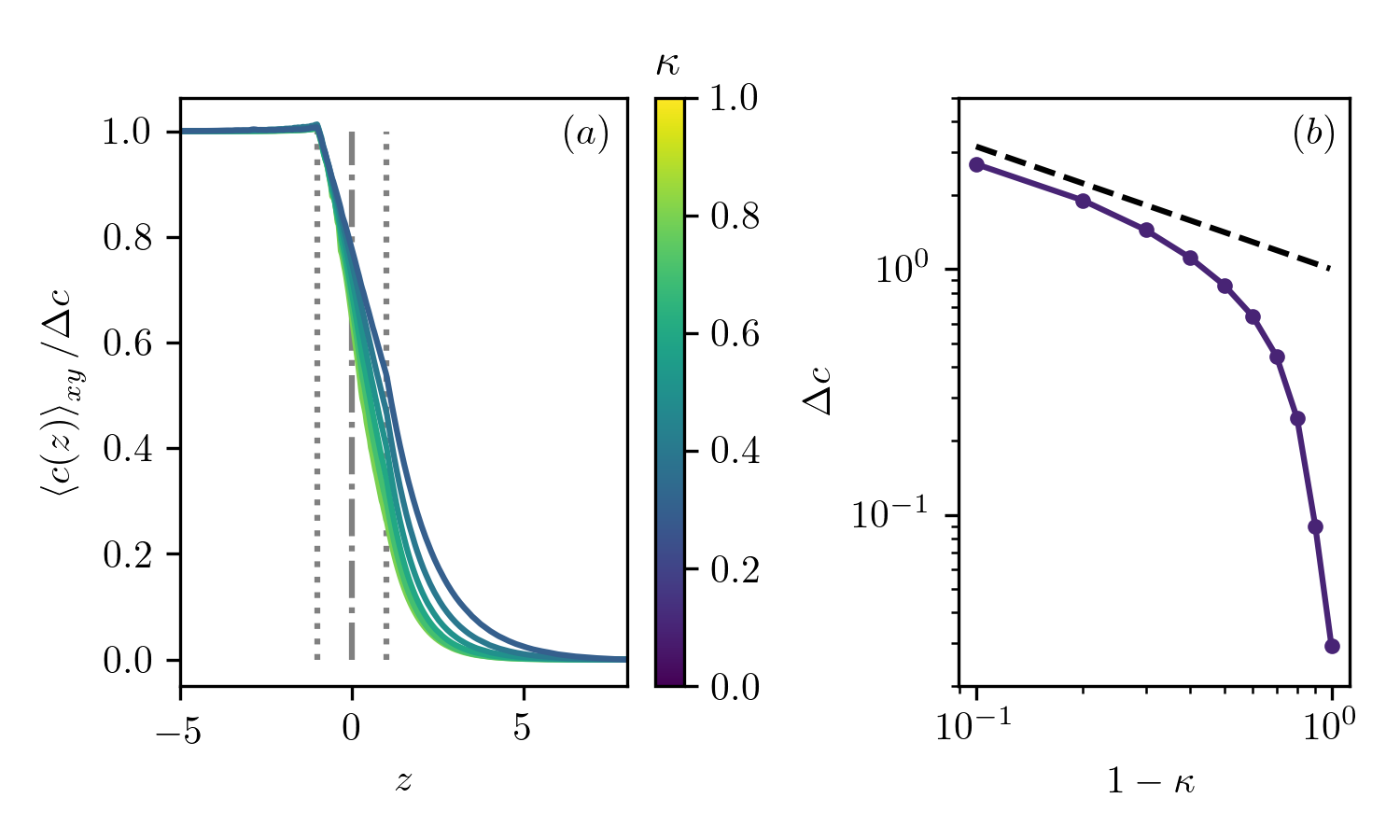}
    \caption{(a) Evolution of the spanwise-averaged solute concentration in the fluid around the particle along the pipe for increasing confinement $\kappa$ (color) for $\Pe=6$. The particle's position and limits are shown (resp. dash-dotted and dotted lines). \sm{The results are reported for $\kappa=n/10$ with $2\leq n\leq 8$.}  (b) Evolution with $\kappa$ of the front-back concentration contrast measured using the spanwise-averaged concentration at $|z|\gg R$ upstream and downstream from the particle. The dashed line corresponds to the analytical prediction of Section~\ref{sec:theory}), see Eq.~\eqref{eq:Upredictions2}.}
    \label{fig:Pipe_Pe_6.0_kappa_f_spanwise}
\end{figure}

The dominance of lateral diffusion, and resulting homogenisation of the concentration in most of the domain (i.e. apart from $|z|\sim 1$), justifies focusing on the mean concentration along the capillary axis, defined as the average within each cross section (fixed $z$):
\begin{equation}
\langle c\rangle _{xy}(z)=\frac{1}{\pi (R^2-R_\textrm{min}(z)^2)}\int_{R_\textrm{min}(z)}^R\int_0^{2\pi}c(r,z)r\dd r\dd \theta
\end{equation}
with $R_\textrm{min}(z)=\sqrt{a^2-z^2}$ for $|z|<a$ and $R_\textrm{min}(z)=0$ otherwise. This function of $z$ only  takes a uniform value far upstream and downstream of the particle (Figure~\ref{fig:Pipe_Pe_6.0_kappa_f_spanwise}), so that the front-back concentration contrast can be defined as 
\begin{equation}
\Delta c=\langle c\rangle _{xy}(z=-L/2)-\langle c\rangle _{xy}(z=L/2).
\end{equation}
 Figure~\ref{fig:Pipe_Pe_6.0_kappa_f_spanwise}(a) shows that the evolution with $z$ of the average concentration, once rescaled by $\Delta c$,  becomes essentially independent of $\kappa$ for $\kappa\gtrsim 0.5$, and that this universal profile is characterised by (i) constant values behind and ahead of the particle ($z<-1$ or $z\gtrsim 2$) and (ii) a linear profile (constant streamwise gradient) in most of the vicinity of the particle.   The amplitude of the front-back concentration contrast increases however sharply with $\kappa$, diverging as $(1-\kappa)^{-1/2}$ as $\kappa\rightarrow 1$, demonstrating the confinement-induced chemical saturation (Figure~\ref{fig:Pipe_Pe_6.0_kappa_f_spanwise}b). This behaviour is quantitatively consistent with the asymptotic predictions, see Eq.~\eqref{eq:Upredictions2} and Section~\ref{sec:theory}. 
 
 \smbis{The fore-aft asymmetry of the concentration profile, observed on Figure~\ref{fig:Pipe_Pe_6.0_kappa_f_spanwise}a for $\Pe=6$, results from the accumulation of solute in the wake of the propelling droplet due to the restricted lateral diffusion when the droplet is sufficiently confined (large enough $\kappa$) and is also present for larger $\Pe$. A small overshoot of the concentration profile can be observed immediately behind the particle for the least confined configurations (small $\kappa$) for which the solute transport balance is fundamentally different.}

\subsection{Effect of confinement on the flow field}

The flow pattern and intensity generated by the swimming particle inside the capillary is also significantly modified by the presence and distance to the neighbouring walls. For strong confinement, the largest fluid velocities and velocity gradients are observed within the thin fluid gap: a finite volume of fluid needs to be moved from one side of the particle to the other through a narrower gap in order to allow for the particle motion through the capillary where the flow is at rest away from the particle. As $\kappa$ approaches $1$, the typical fluid velocity within the gap is therefore much higher than the particle velocity itself (see also Section~\ref{sec:theory} for a more quantitative discussion), resulting in strong spanwise gradients of the fluid streamwise velocity within the narrow gap (Figure~\ref{fig:pipe_concentration_streamplot}). 

A more detailed analysis of the velocity distribution within the fluid gap further reveals that, as $\kappa$ is increased, the velocity profile tends to a Couette flow profile (Figure~\ref{fig:spanwise_gap}b): the dominant fluid transport in the narrowest fluid layer is therefore driven solely by the phoretic slip at the \sm{particle surface}, resulting from the front-back concentration contrast observed in strongly-confined configurations (Figure~\ref{fig:pipe_concentration_streamplot}). In particular, the absence of curvature in the velocity profile indicates that longitudinal pressure gradients play a negligible effect on the dominant flow.

\begin{figure}
    \centering
    \includegraphics[width=1.0\textwidth]{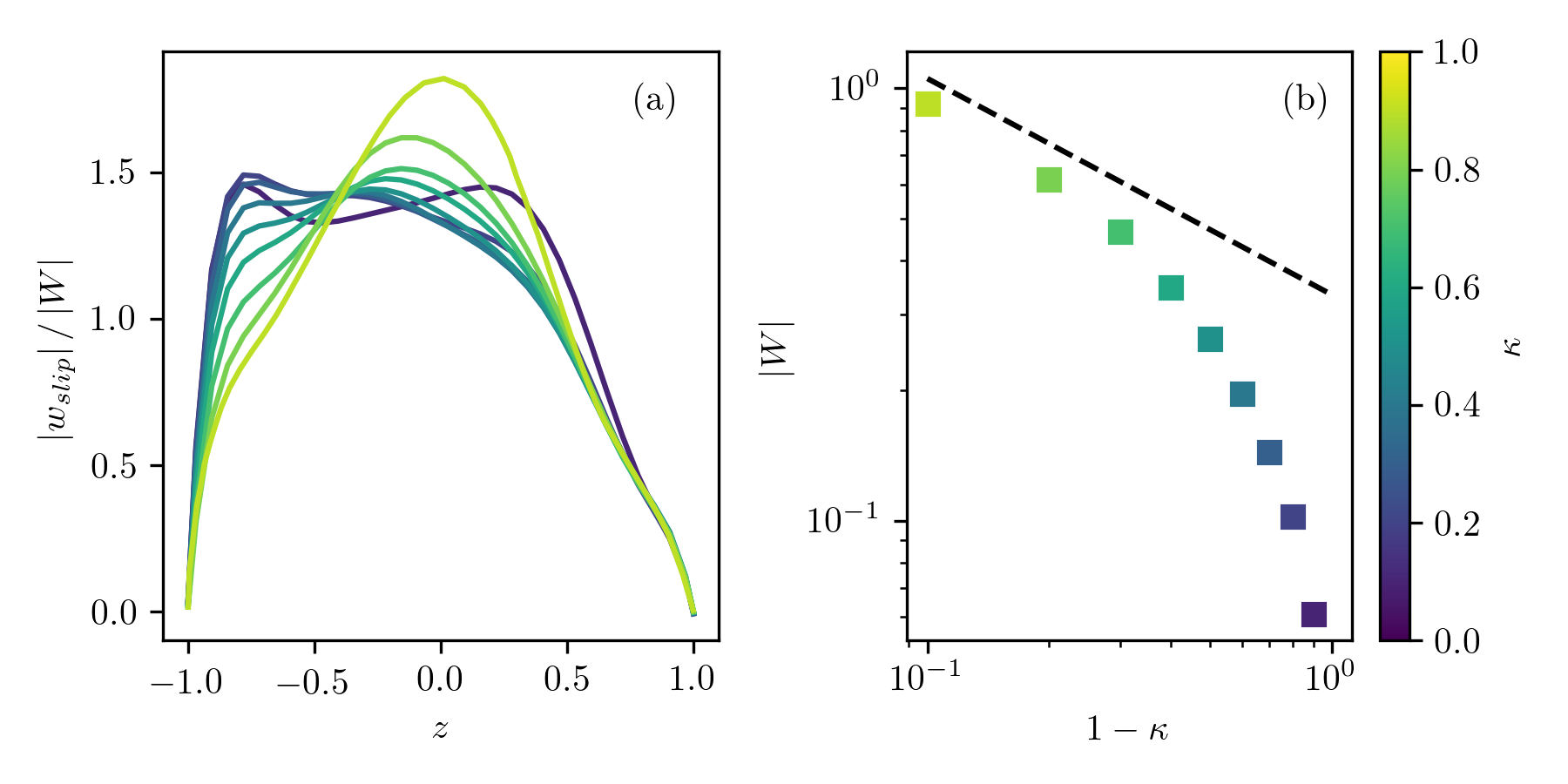}
    \caption{(a) Distribution of the surface slip velocity normalized by the average fluid velocity in the particle-capillary gap. \sm{The results are reported for $\kappa=n/10$ with $2\leq n\leq 8$.}  (b) Average fluid velocity in the particle-capillary gap $W$ at varying confinement $\kappa$. Results are shown here for $\Pe=6.0$. The dashed line corresponds to the analytical prediction, Eq.~\eqref{eq:Upredictions2}. }    
    \label{fig:surface_slip_W}
\end{figure}

We therefore turn our attention to the evolution of this slip forcing for increasing $\kappa$, and more specifically on its streamwise component that plays a major role in the thinnest regions. Once again, a transition can be clearly seen between two different regimes (Figure~\ref{fig:surface_slip_W}a): for weak confinement, the relative distribution of slip is rather constant along the sphere, except near the front and back poles. A slight maximum is observed at the back of the particle, which is in qualitative agreement with the established result that the particle acts as a pusher swimmer in unbounded domains~\citep{Michelin2013,Izri2014}. As confinement is increased, the slip profile becomes more front-back symmetric with a maximum value attained in the narrowest region: this indicates a stronger localisation of the forcing in the regions where it has the most hydrodynamic influence on the self-propulsion. Note that such localisation in the regions of strongest hydrodynamic influence was also recently identified for a chemically-active droplet propelling along a planar wall~\citep{Desai2021}. The evolution of the maximum phoretic slip with $\kappa$ confirms the enhancement of the phoretic forcing as the distance to the confining walls is reduced,\sm{ with the average fluid velocity in the gap, $W=\langle w(z=0)\rangle_{xy}$,} diverging as $(1-\kappa)^{-1}$ when $\kappa\rightarrow 1$. This increase of the phoretic slip with confinement directly results from the increased (and diverging as $\kappa\rightarrow 1$) concentration contrast between the front and back of the phoretic particles that was discussed in greater details in Section~\ref{sec:concentration}. The results are in good agreement with the asymptotic prediction for the evolution of $W$ as $\kappa\rightarrow 1$ (Figure~\ref{fig:surface_slip_W}b and Eq.~\eqref{eq:Upredictions2}).

Finally, the velocity field away from the particle (i.e. upstream and downstream) is almost uniform and eventually decays to zero (in the laboratory reference frame). Here a brief comment should be made regarding the boundary conditions imposed at the inlet and outlet boundaries of the computational domain  (Figure~\ref{fig:sketch}). A Dirichlet boundary condition on the flow velocity is imposed on $\Gamma_\textrm{in}$ and $\Gamma_\textrm{out}$, representing that the flow is at rest far upstream and downstream from the particle in the lab frame. This will be the case for example when the domain considered (Figure~\ref{fig:sketch}) is part of an infinitely long tube: away from the particle, the large hydrodynamic resistance prevents the existence of any flow within the tube.

\subsection{Resistance to particle motion and pressure}

\begin{figure}
    \centering
    \includegraphics[width=1.0\textwidth]{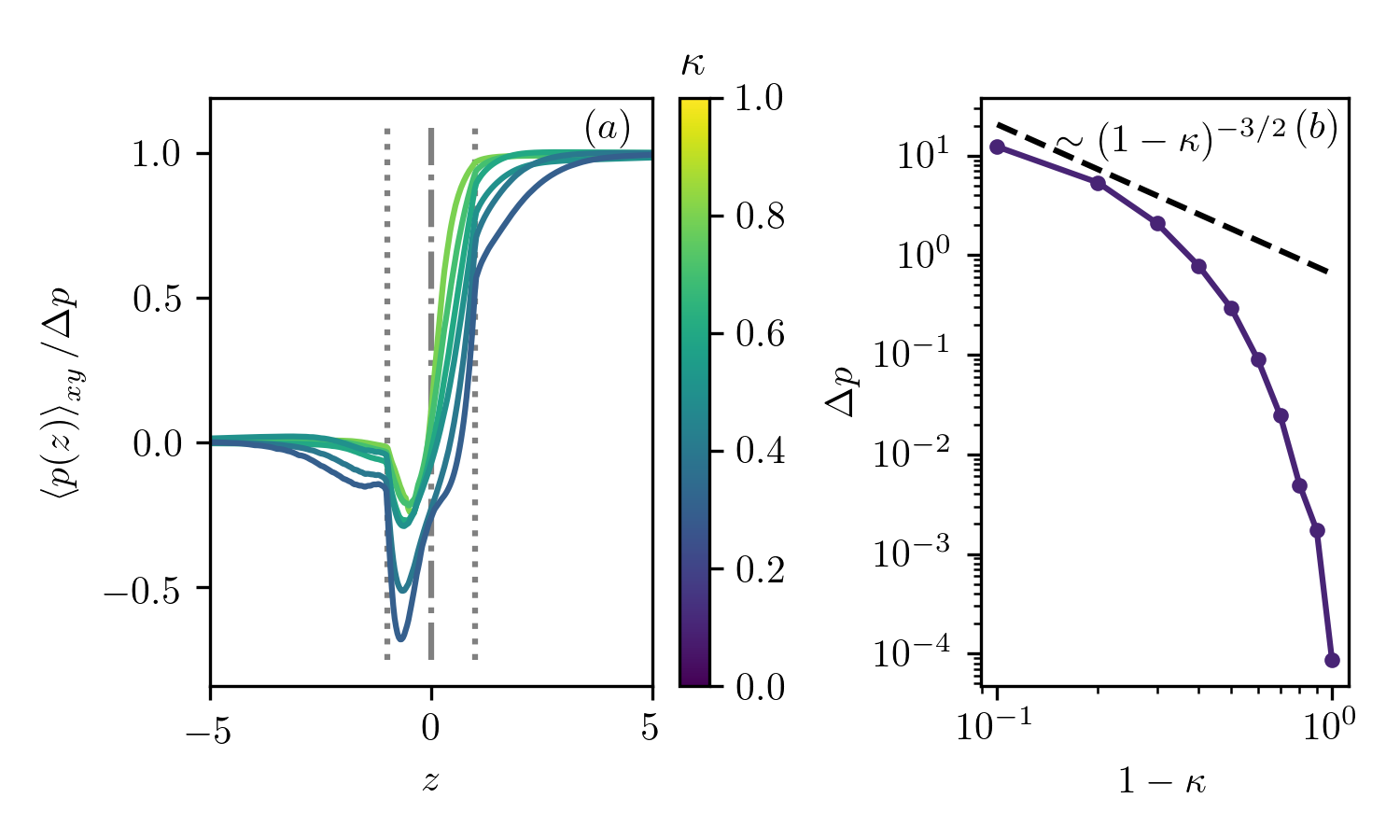}    
\caption{(a) Evolution of the spanwise-averaged fluid pressure in the fluid around the particle for increasing confinement $\kappa$ (color). The particle's position and limits are shown (resp. dash-dotted and dotted lines). \sm{The results are reported for $\kappa=n/10$ with $1\leq n\leq 9$.}  (b) Evolution with $\kappa$ of the front-back mean pressure difference measured using the spanwise-averaged pressure at $|z|\gg R$ upstream and downstream from the particle. In both panels, results were obtained for $\Pe=6$. The dashed line with $(1-\kappa)^{-3/2}$ is shown for comparison only.}
    \label{fig:Pipe_Pe_6.0_kappa_p_spanwise}
\end{figure}

We noted earlier that, because fluid is at rest in front \sm{of } and behind the phoretic particle, a finite volume of fluid must pass through the thin fluid gap for the particle to move forward. Driving such a volume flux through a thin viscous fluid layer \sm{results in the establishment of a net }pressure difference between the front and back of the sphere, as demonstrated on Figure~\ref{fig:Pipe_Pe_6.0_kappa_p_spanwise} by the evolution of the spanwise-averaged pressure $\langle p \rangle_{xy}$ along the capillary (i.e. its average on each cross-section) and of the front-back pressure difference $\Delta p$ (both quantities defined in a similar way as their counterparts for the concentration). 

We first note that the pressure difference $\Delta p$ vanishes when $\kappa\ll 1$, i.e. for unbounded phoretic particles, as expected. When $\kappa$ becomes larger, the pressure still reaches constant values far upstream and downstream of the particle, but they are now different and their difference quickly grows with confinement and diverges as $\kappa\rightarrow 1$ (Figure~\ref{fig:Pipe_Pe_6.0_kappa_p_spanwise}b). While a clear scaling is difficult to identify from the numerical results, it can still be concluded that the divergence observed is weaker than $(1-\kappa)^{-2}$ (see Section~\ref{sec:asymptotics} for further discussion).  
The emergence of a finite (and increasing) pressure difference exerts a resisting force on the particle, balancing the net forcing exerted within the thin fluid gap by the phoretic flows generated by the particle.

It should further be noted that the pressure variations are not monotonic, showing a local minimum in the vicinity of the narrowest regions (Figure~\ref{fig:Pipe_Pe_6.0_kappa_p_spanwise}a).

\section{Self-propulsion of a tightly-fitting particle}\label{sec:theory}
\subsection{\sm{Global conservation arguments}}\label{sec:scaling}

The analyses and results of the previous sections provide some critical insight on the physical balances and phenomena determining the evolution of the confined self-propulsion, in particular in the limit of strong confinement ($\kappa\gtrsim 0.5$).

In the following, these different arguments are summarised and combined to obtain a prediction for the  scaling of the swimming velocity in this limit, in terms of the two main parameters of the problem, $\Pe$ and $\kappa$. Throughout, we focus exclusively on the steady-state regime at the center of our attention in Section~\ref{sec:results}. We will relate three specific quantities: (i) $U_z$ the swimming velocity of the phoretic particles, (ii) $\Delta c$ the difference in the uniform solute concentration observed far downstream and upstream of the particle, respectively, and (iii) $W=\langle w(z=0)\rangle_{xy}$ the mean flow velocity (relative to the particle) through the narrowest fluid region ($z=0$) (oriented along $-\eb_z$, i.e. from the front toward the back).

\subsubsection{Solute conservation}\label{sec:solute_cons}
Considering the entire computational domain as a control volume, the conservation of solute  imposes 
\begin{equation}
\int_{\Gamma}\mathbf{j}\cdot \nb\dd S=0\label{eq:solute_conservation1}
\end{equation}
with $\Gamma=\Gamma_p\cup\Gamma_d\cup\Gamma_\textrm{out}\cup\Gamma_\textrm{in}$ (Figure~\ref{fig:sketch}), $\nb$ the unit normal to $\Gamma$ \sm{pointing \emph{into} the fluid domain}, and \sm{ $\ub$ the fluid velocity in the reference frame of the particle. The non-dimensional solute flux $\mathbf{j}=\Pe c\ub-\nabla c$ (characteristic scale: $|\mathcal{A}|$) includes the contributions of convective transport by the fluid flow and diffusion, respectively.}

\begin{figure}
\begin{center}
\includegraphics[width=.6\textwidth]{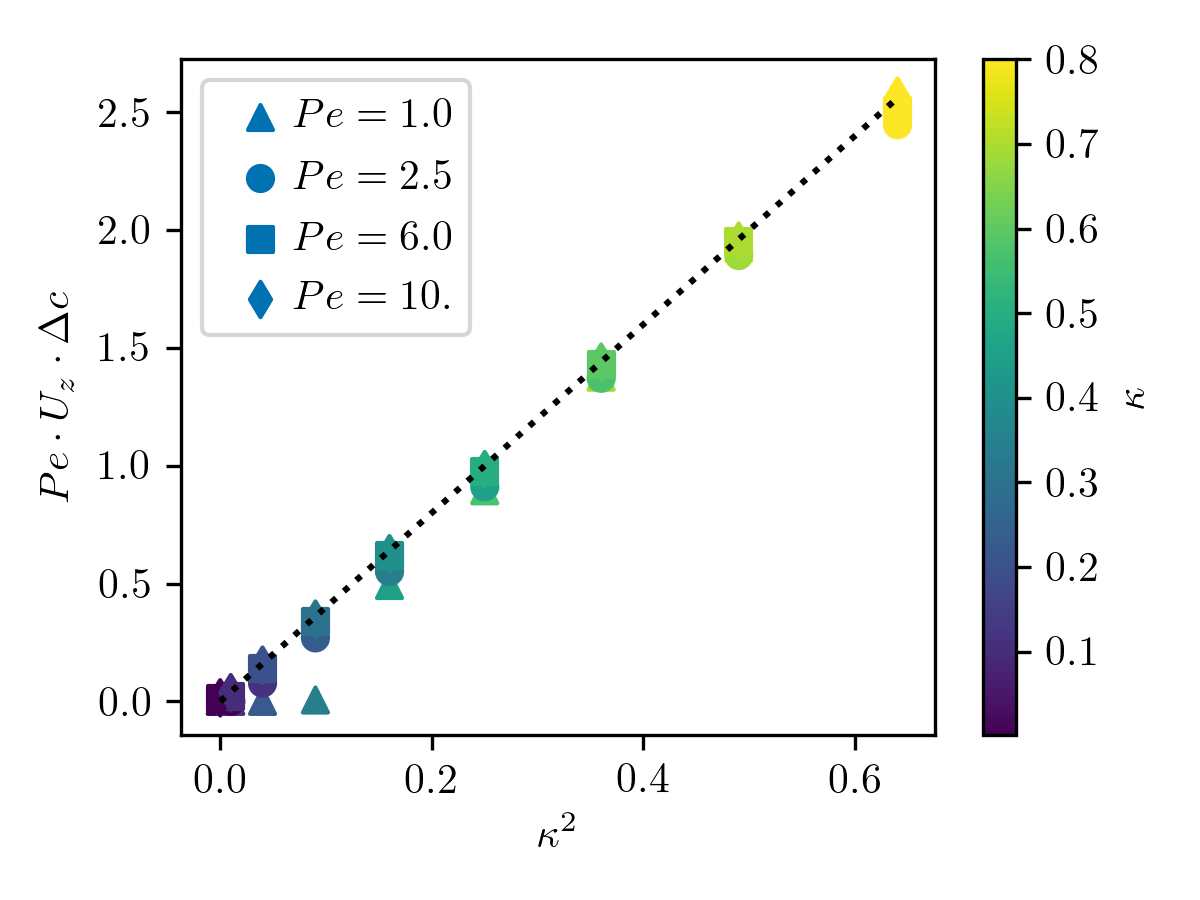}
\caption{\sm{Global conservation of solute: the dimensionless solute emission ($\kappa^2$) is compared to the dimensionless excess solute convected downstream from the sphere ($\Pe\,U_z\Delta c$).} }\label{fig:check_scaling}
\end{center}
\end{figure}

The channel's wall are inactive and impermeable so that $\mathbf{j}\cdot\nb=0$ on $\Gamma_d$. At the particle surface, the solute flux is purely diffusive and matches the total production rate at the particle surface $\int_{\Gamma_p}\mathbf{j}\cdot\nb\dd S=\sm{4\pi}$.
Far from the particle, near $\Gamma_\textrm{in}$ and $\Gamma_\textrm{out}$, the concentration is uniform so that the diffusive flux is negligible on these surfaces. The velocity is also uniform and equal to $-U_z\eb_z$ so that
\begin{equation}
\int_{\Gamma_\textrm{in}\cup\Gamma_\textrm{out}}\mathbf{j}\cdot\nb\dd S=\sm{-}\frac{\pi \sm{\mbox{Pe}\,}U_z\Delta c}{\kappa^2}\cdot
\end{equation}
Equation~\eqref{eq:solute_conservation1} then leads to
\begin{equation}\label{eq:solute_conservation}
\Pe \,U_z\Delta c=4\kappa^2.
\end{equation}
This result was found in agreement with the numerical results for strong enough confinements ($\kappa \gtrsim 0.2$\sm{, Figure~\ref{fig:check_scaling}}). \sm{Note that for lower $\kappa$, such a balance is not expected to hold as the transport mechanism of the solute away from the particle surface vicinity is fundamentally different.}

\subsubsection{Conservation of mass}\label{sec:mass_cons}
A similar argument for the conservation of mass on the upstream half of the computational domain leads to
\begin{equation}
\int_{\Gamma^+}\ub\cdot\nb\dd S=0,
\end{equation}
with $\Gamma^+=\Gamma_\textrm{in}\cup\Gamma_p^+\cup\Gamma_d^+\cup\Gamma_0$ with $\Gamma_d^+$ and $\Gamma_p^+$ the parts of the wall and particle surfaces with $z>0$, and $\Gamma_0$ the fluid cross section at $z=0$. The particle surface $\Gamma_p$ and the wall $\Gamma_d$ are impermeable and do not contribute to the integral above. On $\Gamma_\textrm{in}$, the velocity is uniform and equal to $-U_z$, so that $\int_{\Gamma_{in}}\ub\cdot\nb\dd S=\pi U_z/\kappa^2$. By definition of $W$, $\int_{\Gamma_0}\ub\cdot\nb\dd S=-\pi (1-\kappa^2)W/\kappa^2$, so that
\begin{equation}\label{eq:scaling_mass}
U_z=(1-\kappa^2)W.
\end{equation}

\subsubsection{Fluid velocity trough the gap}
One of the main features of the flow within the thin fluid gap identified in Sec.~\ref{sec:velocity} when the gap thickness is reduced (i.e. $1-\kappa \ll 1$) was the emergence of a Couette-like dominant flow driven by the slip velocity \sm{$w_\textrm{slip}$} at the surface of the particle. 
For a cylindrical Couette flow\sm{~\citep{Leal2007}},
\begin{equation}
W=\frac{w_\textrm{slip}}{2}f(\kappa),\quad \textrm{with   }f(\kappa)=\frac{1}{\log(1/\kappa)}-\frac{2\kappa^2}{1-\kappa^2}
\end{equation}
and $f(\kappa\rightarrow 1)=1$ (plane Couette flow).
The non-dimensional slip velocity is $\totd{c}{z}(z=0)\approx \totd{}{z}\langle c\rangle_{xy}$ since the concentration is uniform across the fluid gap for large enough $\kappa$. The results of Figure~\ref{fig:Pipe_Pe_6.0_kappa_f_spanwise} suggest that the variations of $\langle c\rangle_{xy}$ with $z$ are almost linear so that \sm{the axial concentration gradient in the gap is proportional to the front-back concentration contrast $\Delta c$ and, accordingly, the phoretic slip and mean flow in the gap $W$ satisfy}
\begin{equation}
W\approx K\Delta c\label{eq:WDeltaC}
\end{equation}
\sm{with $K$ a constant of proportionality.}

\begin{figure}
    \centering
    \includegraphics[width=0.75\textwidth]{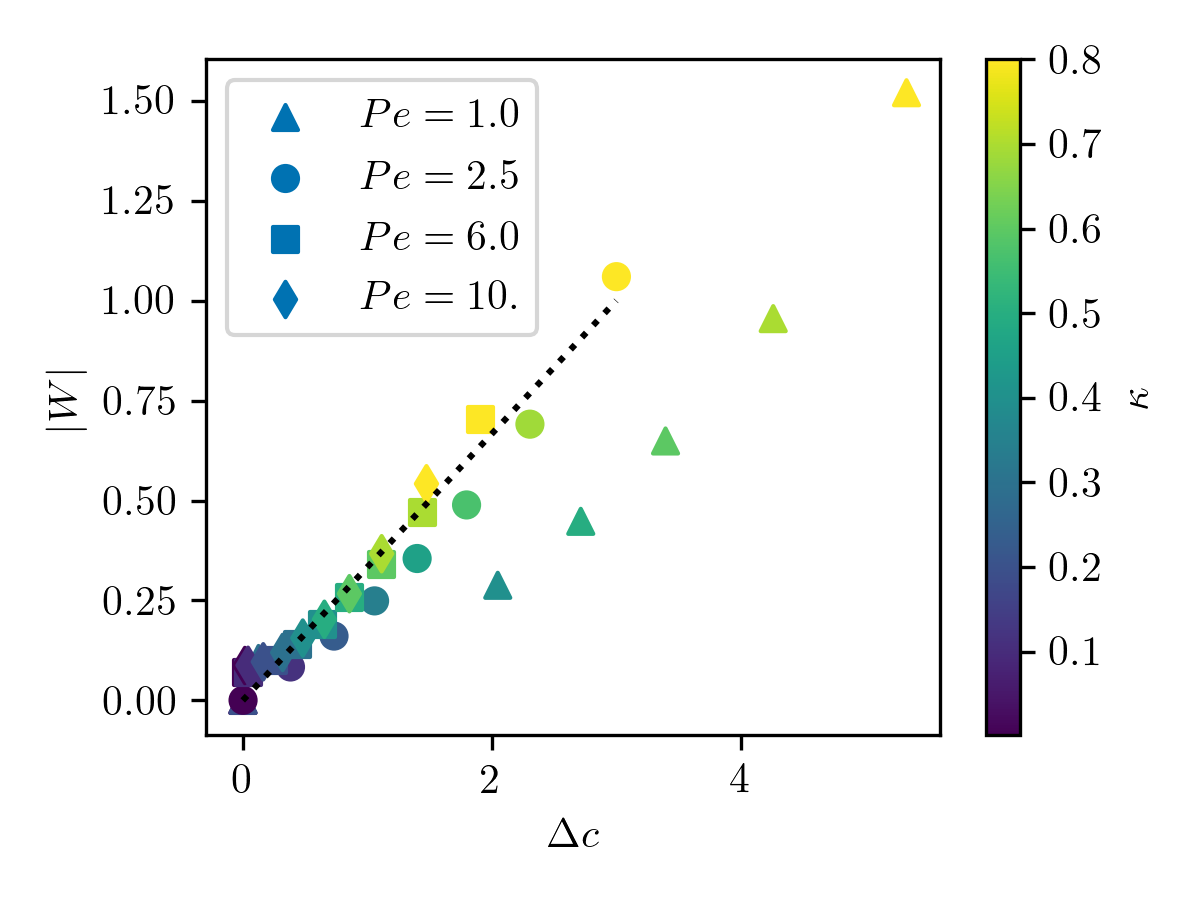}
    \caption{Relative evolution of the spanwise-averaged axial velocity $W$ of the fluid (relative to the particle) at $z=0$ (narrowest fluid gap) and of the front-back solute concentration $\Delta c$. The dotted line corresponds to $\Delta c/3$.}
    \label{fig:W_propto_DC}
\end{figure}
Figure~\ref{fig:W_propto_DC} provides supporting evidence of this linear relationship between the average velocity $W$ and the front-back concentration contrast, with ${K\approx 1/3}$, except for the lowest $\Pe$-values.

\subsubsection{Approximation of the particle velocity}
A combination of macroscopic conservation principles\sm{, Eqs.~\eqref{eq:solute_conservation} and \eqref{eq:scaling_mass},} and qualitative argument\sm{, Eq.~\eqref{eq:WDeltaC},} allowed us to obtain three independent relationships between the three quantities of interest $U_z$, $W$ and $\Delta c$. Combining these provide the following predictions for each of these quantities:
\begin{equation}\label{eq:Upredictions}
U_z\approx 2\kappa\sqrt{\frac{1-\kappa^2}{3\Pe}},\qquad W\approx \frac{2\kappa}{\sqrt{3\Pe(1-\kappa^2)}},\qquad \Delta c \approx 2\kappa\sqrt{\frac{3}{\Pe(1-\kappa^2)}}\cdot 
\end{equation}

\begin{figure}
    \centering
    \includegraphics[width=1.0 \textwidth]{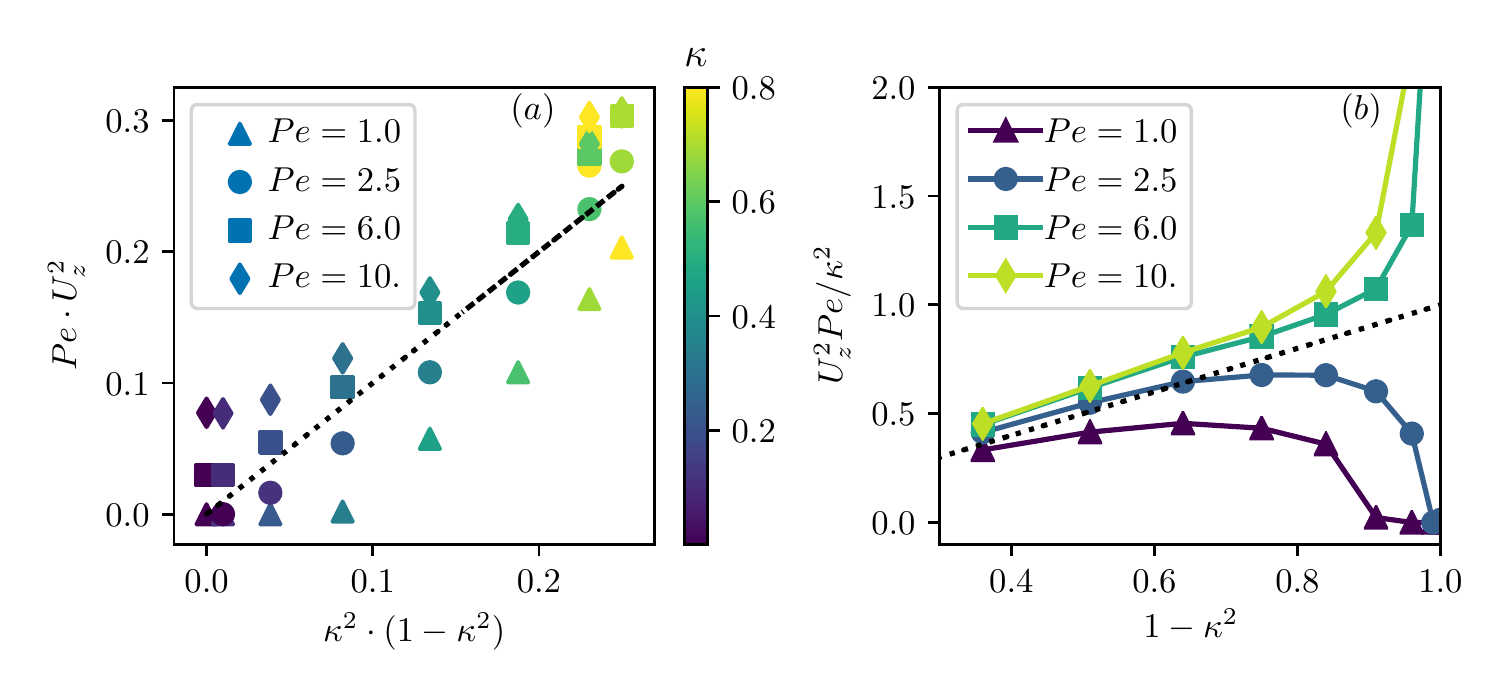}
    \caption{(a) Evolution of the rescaled steady state swimming velocity $U_z$ with $\Pe$ and $\kappa$. In each plot, the dotted line corresponds to the equality of the two plotted quantities. \sm{(b) The same data is reported to identify the behaviour for $\kappa\rightarrow 1$.}
    }
    \label{fig:semi-analytic-scaling}
\end{figure}
These predictions are in quantitative agreements with the numerical results (Figure~\ref{fig:semi-analytic-scaling}) in particular for larger $\kappa$ (i.e. $\kappa\gtrsim 0.5$ or $1-\kappa^2\lesssim 0.7$), except for the lowest value of $\Pe$ investigated. This better agreement for larger $\Pe$ was to be expected as convective transport of solute plays a dominant role in that limit.

\begin{figure}
    \centering
    \includegraphics[width=0.75\textwidth]{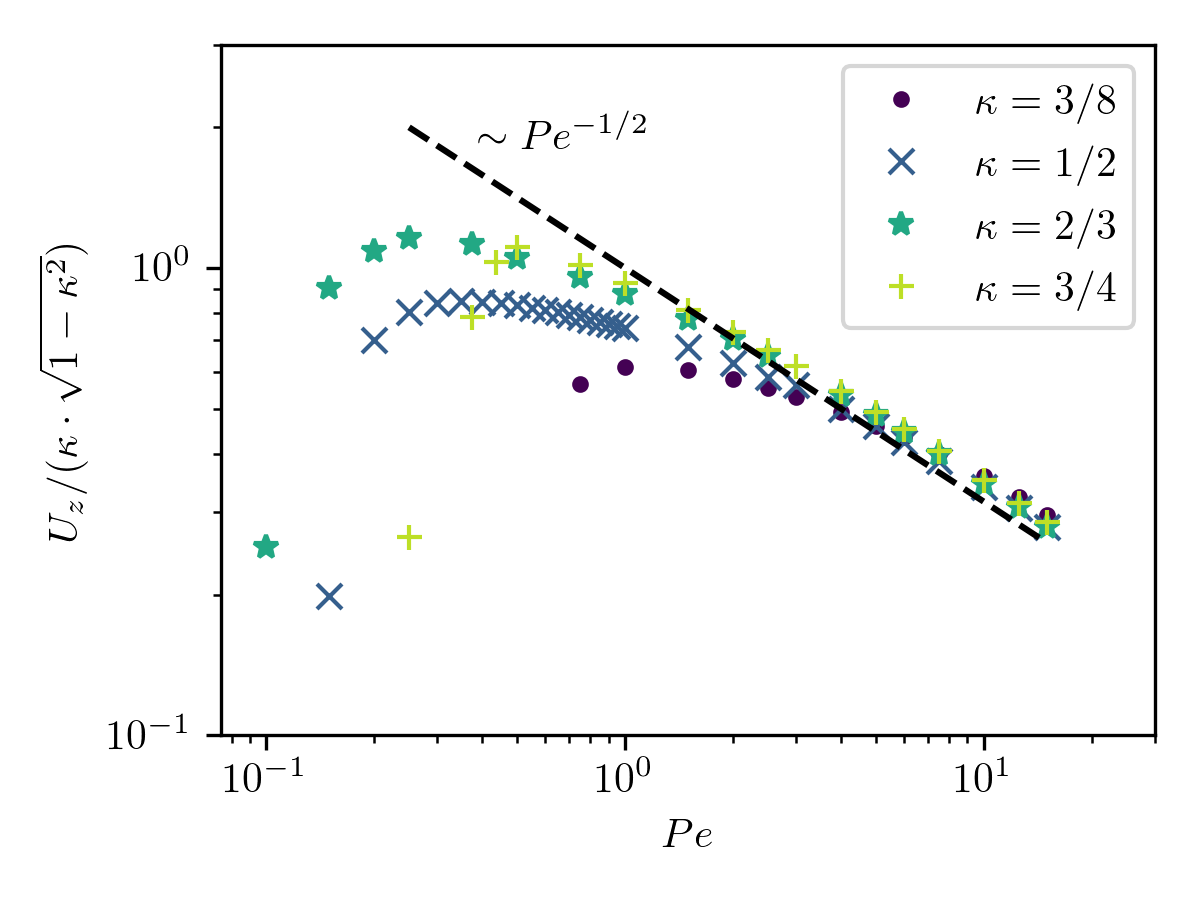}
    \caption{Evolution of the rescaled particle  velocity with $\Pe$ for different confinement levels, $\kappa$. The dashed line corresponds to $\Pe^{-1/2}$. }
    \label{fig:U_Pe_kappa_Law}
\end{figure}

Furthermore, this relationship shows that $U_z$ is not a monotonic function of $\kappa$ but instead should be maximum around $\kappa=1/\sqrt{2}\approx 0.7$ in agreement with the results of Figure~\ref{fig:Pe_critic_kappa}.
These predictions also clearly establish that $U_z\sim 1/\sqrt{\Pe}$, in particular for larger $\Pe$ and larger $\kappa$, which is confirmed in Figure~\ref{fig:U_Pe_kappa_Law}.  
We finally observe that the numerical evolution of $W$ and $\Delta c$ with $\kappa$ are consistent with these predictions (see Figures~\ref{fig:Pipe_Pe_6.0_kappa_f_spanwise} and \ref{fig:surface_slip_W}). 

\subsection{Asymptotic analysis}
\label{sec:asymptotics}
We focus now specifically on the lubrication limit, i.e. when $R\approx a$ or equivalently $\kappa\rightarrow 1$ and thus define $\varepsilon=1-\kappa\ll 1$. Note that, the result above establishes that the dominant swimming velocity is set solely by the slip forcing inside the hydrodynamic lubrication region of width $\sqrt{\varepsilon}$ around the region of smallest thickness (and not anywhere else). In turn, this requires knowing the leading order evolution of the surface concentration in that region.

The scalings obtained from the balance arguments of the previous sections indicate that \sm{for $\varepsilon=1-\kappa\ll 1$,}
\begin{equation}
U_z=\sm{O\left(\sqrt{\frac{\varepsilon}{\Pe}}\right)}\quad\textrm{and  }\quad \sm{W,\,\Delta c=O\left(\frac{1}{\sqrt{\varepsilon\Pe}}\right)}.
\end{equation}
Physically, this indicates that as the fluid layer between the particle and the wall is reduced, the velocity of the particle tends to zero while the front-back concentration difference diverges. This is not surprising as we focus here on the \emph{steady} self-propulsion of the particle. In Section~\ref{sec:concentration}, we noted that the confined limit of the particle self-propulsion corresponds to a fundamental change in the way the solute produced at the particle surface is evacuated: when $\kappa\rightarrow 0$ (unbounded flow), the solute is mostly diffused away in the far field and in all directions, while for $\kappa\rightarrow 1$, it must be convected downstream by the displacement of the particle, as steady diffusive solutions do not exist for these confined configurations. Lower self-propulsion velocities (e.g. due to the increase of viscous stresses at the boundary) therefore require larger concentration accumulation in the back of the self-propelled particle.

These arguments demonstrate not only a typical $O(\varepsilon^{-1/2})$-scale for the magnitude of the concentration in the thin lubrication layer located between the particle and the wall (with respect to the reference concentration far ahead of the particle, taken as zero here), but also that this concentration contrast $\Delta c$ is established at the scale of the size of the particle, so that the relevant scale of horizontal variations for $c$ is $O(1)$ not the classical $O(\varepsilon^{1/2})$-length of the  lubrication zone relevant for hydrodynamic lubrication problems. Within the thin fluid layer surrounding $z=0$, one must therefore expect 
\begin{equation}
c=O(\varepsilon^{-1/2}),\qquad \tilde{u}\sim\pard{c}{x}=O(\varepsilon^{-1/2}),
\end{equation}
where $\tilde u$ is the slip velocity forcing at the \sm{particle surface}.

\subsubsection{\sm{Hydrodynamic lubrication}}
When $\varepsilon\rightarrow 0$, the fluid's motion within the thin annular layer surrounding the particle at $z=0$ corresponds to a lubrication flow forced at the surface of the particle by the phoretic slip $\tilde{u}(z)$ along its surface. It is therefore described by the two-dimensional lubrication equations
\begin{equation}
\pard{p}{\rho}=0,\quad \pard{^2\sm{u_z}}{\rho^2}=\pard{p}{z},\quad\pard{u_z}{z}+\pard{u_\rho}{\rho}=0,
\end{equation}
with $\rho=\kappa^{-1}-r$ the radial distance from the outer cylinder of radius $1/\kappa$ (measured inward), and $u_\rho$ the corresponding velocity component. The  boundary conditions on the axial velocity are at leading order
\begin{equation}
u_z(\rho=0,z)=-U_z,\qquad u_z(\rho=h(z),z)=\tilde{u}(z),
\end{equation}
with $\rho=h(z)$ the equation for the surface of the particle, i.e.
\begin{equation}
h(z)=\frac{1}{\kappa}-\sqrt{1-z^2}\sim\sm{\varepsilon}\left[1+\frac{z^2}{2\varepsilon}+O\left(\frac{z^4}{\varepsilon}\right)\right].
\end{equation}
\sm{Note that the present analysis is similar to that developed for the electrophoretic motion of a sphere inside a tightly-fitting tube~\cite{Sherwood2018}.} 
The lubrication equations can be integrated to find the axial flow $u_z(z)$,
\begin{equation}
u_z(\rho,z)=\frac{\rho(\rho-h)}{2}\sm{\totd{p}{z}}+\tilde{u}(z)\left(\frac{\rho}{h}\right)+U_z\left(\frac{\rho}{h}-1\right),
\end{equation}
and integration across the fluid layer and around the particle provides the volume flux 
\begin{equation}\label{eq:reynolds}
q(z)=Q=2\pi\int_0^hu_z(\rho,z)\dd \rho=-\frac{\pi h^3}{6}\sm{\totd{p}{z}}+\pi h\left(\sm{-}U_z+\tilde{u}(z)\right),
\end{equation}
which must indeed be a constant for all $z$ in order to conserve the total volume flux through the different cross sections. We first note that $Q=2\pi \varepsilon W$, with $W$ the average axial velocity within the narrowest gap. 
Using the conservation of mass around the particle, Eq.~\eqref{eq:scaling_mass}, as $\varepsilon\rightarrow 0$, we further note that 
\begin{equation}
Q=2\pi\varepsilon W=\pi U_z \label{eq:lub_Q}
\end{equation}
when $\varepsilon\rightarrow 0$ so that the $U_z$-contribution to the right-hand side of Eq.~\eqref{eq:reynolds} \sm{is $O(\varepsilon U_z)$ and is negligible in front of the left hand side of that equation, as $Q=O(U_z)$}.

Classically, this equation can then be used to compute the pressure difference between the two ends of the hydrodynamic lubrication region~\citep{Leal2007}
\begin{equation}
\Delta P=\int_{-l}^{l}\pard{p}{z}\dd z=-6U_z\int_{-l}^l\frac{\dd z}{h^3}+6\int_{-l}^l\frac{\tilde{u}(z)\dd z}{h^2},\label{eq:lub_DeltaP}
\end{equation}
with \sm{$l\gg \varepsilon^{1/2}$ a length scale much larger than the typical $\varepsilon^{1/2}$-width of the lubrication region}. We note that because $1/h$ varies from $0$ to $1/\varepsilon$ over a \sm{ $O(\varepsilon^{1/2})$ length scale}, the \sm{two integrals on the right-hand side of Eq.~\eqref{eq:lub_DeltaP}} scale respectively as $O(U_z\varepsilon^{-5/2})$ and $O(\tilde{u}\varepsilon^{-3/2})$.

\sm{The phoretic slip $\tilde{u}$, as the forcing phenomenon of the problem, should remain part of the dominant balance in the conservation of volume flux, Eq.~\eqref{eq:reynolds}, so that $Q=O(\varepsilon \tilde{u})$. As a result, and using Eq.~\eqref{eq:lub_Q}, both terms on the right hand side of Eq.~\eqref{eq:lub_DeltaP} are of the same order and contribute to the dominant balance.}

\sm{The left-hand side of Eq.~\eqref{eq:lub_DeltaP} represents a pressure difference between the upstream and downstream regions \sm{away from the particle. This} would lead to a $O(\Delta P)$ resistive force on the phoretic particle, \sm{that} must be balanced by a driving force of the same order \sm{for self-propulsion to occur}. This driving force can only arise from the phoretic slip forcing and associated shear stress $\pard{u_z}{\rho}=O(\tilde{u} \varepsilon^{-1})$ at the particle's boundary within the lubrication zone, resulting in a $O(\tilde{u}\varepsilon^{-1/2})$-driving force on the particle once integrated over the $O(\varepsilon^{1/2})$-lubrication region, so that, at most $\Delta P=O(\tilde{u}\varepsilon^{-1/2})$.  }
This establishes that $\Delta P$ is subdominant in Eq.~\eqref{eq:lub_DeltaP} and both integrals \sm{on the right-hand side of Eq.~\eqref{eq:lub_DeltaP}} must therefore balance exactly. Consequently, the swimming velocity $U_z$, is obtained from the slip velocity $\tilde{u}$ along the \sm{particle surface} as
\begin{equation}
U_z=\frac{\displaystyle\int_{-l}^l\frac{\tilde{u}(z)\dd z}{h^2}}{\displaystyle\int_{-l}^l\frac{\dd z}{h^3}},\label{eq:lub_uswim}
\end{equation}
that demonstrates that $U_z=O(\varepsilon\tilde{u})$.

\subsubsection{Chemical transport through the hydrodynamic lubrication layer}
Note that, the result above establishes that the dominant swimming velocity is set solely by the slip forcing inside the hydrodynamic lubrication region of width $\sqrt{\varepsilon}$ around the region of smallest thickness (and not anywhere else). In turn, this requires knowing the leading order evolution of the surface concentration in that region. It was however noted earlier that the chemical transport is externally constrained by the front-back concentration contrast imposed by the displacement of the particle in a confined setting, so that axial concentration gradients are essentially constant along the hydrodynamic lubrication region.

As for the hydrodynamic lubrication, the leading order problem for the concentration is two-dimensional in the $(z,\rho)$-plane. 
The relevant boundary conditions satisfied by the concentration field are then
\begin{align}
\pard{c}{\rho}&=0\quad\textrm{at  }\rho=0,\label{eq:lubc_bc1}\\
\pard{c}{\rho}&=1+h'\pard{c}{z}\quad\textrm{at  }\rho=h(z),\label{eq:lubc_bc2}
\end{align}
since $\nb=-\eb_\rho+h'\eb_z$ at leading order.

Equations~\eqref{eq:lubc_bc1}--\eqref{eq:lubc_bc2} indicate that the relevant length scale for the variations of $c$ in the $\rho$ direction is $h=O(\varepsilon)$. We noted earlier however that the relevant length scale in the axial $z$-direction is $O(1)$. From the hydrodynamic lubrication problem, we also obtained that $u_z=O(\tilde u)=O(\varepsilon^{-1/2})$ so that by mass conservation $u_\rho=O(1)$. \sm{Furthermore, the steady advection-diffusion equation satisfied by $c$, i.e. ${\Pe\, \ub\cdot\nabla c=\nabla^2c}$, becomes at leading order $\pard{^2c}{\rho^2}=0$, establishing }that the leading order $O(\varepsilon^{-1/2})$ concentration field must necessarily satisfy $\pard{c}{\rho}=0$ and that the non-homogeneous boundary condition, Eq.~\eqref{eq:lubc_bc2}, corresponds to subdominant corrections of the concentration field. 

Since the problem is axisymmetric around the particle, the conservation of solute in the fluid volume contained between two successive cross sections at $z$ and $z+\dd z$ provides the following simplified equation for the evolution of $c(z)$:
\begin{align}
\totd{}{z}\left(h\totd{c}{z}\right)-\frac{\Pe\, Q}{2\pi}\totd{c}{z}+1=0
\end{align}
where the successive terms in the previous equation arise from the balance of diffusion, convection by the flow within the lubrication layer and production at the particle surface, respectively. 

Previously, we noted that the variations of $c$ along the $z$-direction occur at the \sm{$O(1)$} scale of the particle. The dominant transport balancing the production at the particle surface is therefore purely convective (diffusive terms are subdominant in the hydrodynamic lubrication region). 
As a result the leading order axial concentration gradient, and surface slip velocity, are constant and obtained simply as 
\begin{align}
\tilde u=\pard{c}{z}=\frac{2}{\sm{\Pe \,U_z}}.
\end{align}
Reporting this result into Eq.~\eqref{eq:lub_uswim}, we obtain
\begin{equation}
\sm{\Pe \,U_z^2}=\frac{2\displaystyle\int_{-l}^l\frac{\dd z}{h^2}}{\displaystyle\int_{-l}^l\frac{\dd z}{h^3}}\cdot\label{eq:int_frac}
\end{equation}
The integral at the numerator can be expanded as follows, keeping only the leading order contribution as $\varepsilon\ll 1$ and $l\gg \sqrt{\varepsilon}$
\begin{align}
\displaystyle\int_{-l}^l\frac{\dd z}{h^2}=\frac{1}{\varepsilon^2}\displaystyle\int_{-l}^l\frac{\dd z}{\left[1+\sm{z^2/(2\varepsilon)}\right]^2}=\frac{\sqrt{2}}{\varepsilon^{3/2}}\displaystyle\int_{-l/\sqrt{\varepsilon}}^{l/\sqrt{\varepsilon}}\frac{\dd u}{(1+u^2)^2}=\frac{\pi}{\varepsilon^{3/2}\sqrt{2}}\cdot
\end{align}
Similarly, the denominator integral \sm{in Eq.~\eqref{eq:int_frac}} is obtained as $\frac{3\pi}{4\varepsilon^{5/2}\sqrt{2}}$ so that finally,
\begin{align}
U_z=\sqrt{\frac{8\varepsilon}{3\Pe}}\cdot
\end{align}

This result is consistent with the qualitative and quantitative simulations and analysis of Sections~\ref{sec:results} and \ref{sec:scaling} when $\kappa\rightarrow 1$.  It further validates analytically the numerical prefactors obtained in Section~\ref{sec:scaling} from the simulation results for the swimming velocity $U_z$, mean fluid velocity in the gap $W$ and concentration contrast $\Delta c$, Eq.~\eqref{eq:Upredictions} so that as $\kappa\rightarrow 1$, the leading-order behaviour of the swimming velocity, mean fluid velocity in the gap and global concentration contrast are
\begin{equation}\label{eq:Upredictions2}
U_z\sim 2\sqrt{\frac{2(1-\kappa)}{3\Pe}},\qquad W\sim \sqrt{\frac{2}{3\Pe(1-\kappa)}},\qquad \Delta c \sim \sqrt{\frac{6}{\Pe(1-\kappa)}}\sm{\cdot}
\end{equation}

\section{Conclusions and perspectives}\label{sec:conclusions}
Following recent experimental observations and characterization of the behaviour of chemically-active droplets inside small capillaries~\citep{deBlois2021}, the influence of spatial confinement on the self-propulsion of such droplets was investigated here using a combination of direct numerical simulations and asymptotic analysis.

\sm{To overcome the triple challenge posed by the complex geometry of the problem, the nonlinear hydrochemical coupling and the need for a precise implementation of surface boundary conditions, we specifically developed  a novel approach based on embedded boundaries and implemented on top of the popular flow solver Basilisk~\citep{Popinet2015}. Our focus was here on the axisymmetric motion of a single particle along the centerline of a straight capillary. Nevertheless, the framework is} completely general and can be easily adapted to account for more complex geometric domains and/or larger numbers of particles.

Using this versatile numerical tool, we analysed \remove{on the self-propulsion dynamics of a single isotropic phoretic colloid along the central axis of a cylindrical capillary, more specifically} the dual effect of spatial confinement  and of convective transport of solute.  \sm{The particle-to-capillary size ratio, $\kappa$, was} found to  alter significantly the dynamics of an isotropic autophoretic swimmer, generally promoting and enhancing self-propulsion. Indeed, \sm{with increased confinement (larger $\kappa$)}, the \sm{self-propulsion} threshold $\Pe_c$ is starkly reduced, becoming essentially negligible as $\kappa\rightarrow 1$. Additionally, for fixed $\Pe\geq \Pe_c$, the swimmer's velocity \sm{increases significantly} with confinement, up to a maximum value reached for $0<\kappa<1$, before decreasing again and vanishing as $\sqrt{1-\kappa}$ when $\kappa\rightarrow 1$ (near-contact limit). \sm{For fixed $\kappa$}, the swimming velocity  of strongly-confined particles scales as $1/\sqrt{\Pe}$.
 
 Below the self-propulsion threshold $\Pe_c$, convective transport is not sufficient to destabilise a symmetric solution and the system relaxes (in time) toward a front-back symmetric solute distribution and no particle motion. Note that, no steady state can be reached for the concentration whose average value around the particle increases linearly in time as diffusion\sm{, restricted to occur along the axis of the capillary,} is not sufficient to transport away the solute produced at the \sm{particle surface}. Yet, a steady regime is reached for the concentration gradients and flow fields.

These observations \sm{stem from a fundamental alteration of the chemical transport, as the} presence of the confining passive walls prevents solute diffusion away from the particle, except along the capillary axis. Then, convective transport becomes the predominant mechanism to balance the solute production \sm{by the swimmer, resulting in an increased front-aft concentration contrast as the particle leaves a solute-saturated wake behind. }
The phoretic surface slip velocities \sm{are thus increased promoting} the particle's self-propulsion, despite the increased viscous stresses \sm{introduced by} the lateral confinement. 
When $\kappa\rightarrow 1$, the particle dynamics is in fact completely driven by the most confined regions consisting of a thin fluid gap around the particle's equator. The solute distribution is homogeneous across \sm{this thin fluid layer}, and the flow field is completely driven by the phoretic slip at the \sm{particle surface}. 

We confirmed these results in the near-contact limit ($\kappa\rightarrow 1$) using lubrication analysis, that demonstrated that, for $\mbox{Pe}=O(1)$, the concentration gradient inside the lubrication region is, in fact, essentially uniform and set by \sm{the balance of mass and solute between the upstream and downstream regions}, in stark contrast with what is observed for weaker confinement such as a particle near an infinite planar wall~\citep{Desai2021}.
Using these arguments, a predictive model for the swimming velocity \sm{with no fitting parameter was obtained and validated against the direct numerical simulations. This model} confirmed the dependence with $\kappa$ and $\Pe$ of the swimming velocity, as well the existence of an optimal confinement maximising self-propulsion. The detailed dynamics of the solute and particle near the propulsion threshold \sm{($\mbox{Pe}\approx\mbox{Pe}_c$)} remains however to be characterised. 
 
 Throughout this work, we adopted a simplified phoretic model with a rigid particle generating slip flows in response to chemical gradients; yet, the similarity in the solute transport dynamics between rigid isotropic particles and active droplets~\citep{Izri2014,Morozov2019b}, suggests that much of the qualitative conclusions presented here  remain valid if a more complete hydrodynamic description of the droplet is retained, in particular the dominant dependence of the velocity with $(\kappa,\Pe)$. \sm{Despite our focus on a strictly confined geometry (i.e. a capillary surrounding the particle tightly a), our results shed fundamental light on the role of the lubrication layer. This is critical for understanding the propulsion of active droplets along flat boundaries or in Hele-Shaw geometries, although the absence of confining walls around most of the particle surface introduces  key distinctive features in the solute transport and associated dynamics~\citep[e.g. the critical threshold $\Pe_c$ is reduced to a reduced $O(1)$ value as the particle gets closer to the wall, see][]{Desai2021}}.

Finally, our numerical framework unlocks the possibility to simulate the full non-linear hydrodynamic coupling leading to spontaneous motion of autophoretic swimmers under any generic confinement \sm{and for many particles}. In particular, it can be used to analyse the off-axis self-propulsion of the particle and the detailed stability of the axisymmetric solution considered here with respect to fully-$3D$ perturbations, which was purposely left here for a later publication for clarity. This would provide some critical insight into the non-straight motion observed experimentally for mildly-confined active droplets~\citep{deBlois2021}. \sm{It could also provide a much-needed understanding of the individual dynamics of active droplets in complex geometries~\citep{Jin2019} or their collective organisation~\citep{Illien2020,Hokmabad2020b}.}

\begin{acknowledgements}
This work was supported by the European Research Council (ERC) under the European Union’s
Horizon 2020 research and innovation program (Grant Agreement No. 714027 to S.M.). 
\end{acknowledgements}


\begin{thebibliography}{10}

\bibitem{Nelson2010}
B.~J. Nelson, I.~K. Kaliakatsos, and J.~J. Abbott.
\newblock Microrobots for minimally invasive medicine.
\newblock {\em Ann. Rev. Biomed. Eng.}, 12(1):55--85, July 2010.

\bibitem{Dreyfus2005}
R.~Dreyfus, J.~Baudry, M.~L. Roper, M.~Fermigier, H.~A. Stone, and J.~Bibette.
\newblock Microscopic artificial swimmers.
\newblock {\em Nature}, 437(7060):862--865, October 2005.

\bibitem{Magdanz2020}
V.~Magdanz, I.~S.~M. Khalil, J.~Simmchen, G.~P. Furtado, S.~Mohanty,
  J.~Gebauer, H.~Xu, A.~Klingner, A.~Aziz, M.~Medina-S{\'{a}}nchez, O.~G.
  Schmidt, and S.~Misra.
\newblock {IRONSperm}: Sperm-templated soft magnetic microrobots.
\newblock {\em Science Adv.}, 6(28):eaba5855, July 2020.

\bibitem{Lauga2009}
E.~Lauga and T.~R. Powers.
\newblock The hydrodynamics of swimming microorganisms.
\newblock {\em Rep. Prog. Phys.}, 72(9):096601, August 2009.

\bibitem{Purcell1977}
E.~M. Purcell.
\newblock Life at low {R}eynolds number.
\newblock {\em Am. J. Phys.}, 45(1):3--11, January 1977.

\bibitem{Brooks2020}
A.M. Brooks and M.~S. Strano.
\newblock A conceptual advance that gives microrobots legs.
\newblock {\em Nature}, 584(7822):530--531, August 2020.

\bibitem{Bunea2019}
A.-I. Bunea and J.~Gl\"{u}ckstad.
\newblock Strategies for optical trapping in biological samples: Aiming at
  microrobotic surgeons.
\newblock {\em Laser Photonics Rev.}, 13(4):1800227, February 2019.

\bibitem{Koleoso2020}
M.~Koleoso, X.~Feng, Y.~Xue, Q.~Li, T.~Munshi, and X.~Chen.
\newblock Micro/nanoscale magnetic robots for biomedical applications.
\newblock {\em Mat. Today Bio}, 8:100085, September 2020.

\bibitem{Rao2015}
K.~J. Rao, F.~Li, L.~Meng, H.~Zheng, F.~Cai, and W.~Wang.
\newblock A force to be reckoned with: A review of synthetic microswimmers
  powered by ultrasound.
\newblock {\em Small}, 11(24):2836--2846, April 2015.

\bibitem{Moran2019}
J.~Moran and J.~Posner.
\newblock Microswimmers with no moving parts.
\newblock {\em Phys. Today}, 72(5):44--50, May 2019.

\bibitem{Berg1993}
H.C. Berg.
\newblock {\em Random Walks in Biology}.
\newblock Princeton University Press, 1993.

\bibitem{Anderson1989}
J.~L. Anderson.
\newblock Colloid transport by interfacial forces.
\newblock {\em Ann. Rev. Fluid Mech.}, 21(1):61--99, January 1989.

\bibitem{Bechinger2016}
C.~Bechinger, R.~Di Leonardo, H.~L\"{o}wen, C.~Reichhardt, G.~Volpe, and
  G.~Volpe.
\newblock Active particles in complex and crowded environments.
\newblock {\em Rev. Mod. Phys.}, 88:045006, November 2016.

\bibitem{Marchetti2013}
M.~C. Marchetti, J.~F. Joanny, S.~Ramaswamy, T.~B. Liverpool, J.~Prost, Madan
  Rao, and R.~Aditi Simha.
\newblock Hydrodynamics of soft active matter.
\newblock {\em Rev. Mod. Phys.}, 85(3):1143--1189, July 2013.

\bibitem{Howse2007}
J.~R. Howse, R.~A.~L. Jones, A.~J. Ryan, T.~Gough, R.~Vafabakhsh, and
  R.~Golestanian.
\newblock Self-motile colloidal particles: From directed propulsion to random
  walk.
\newblock {\em Phys. Rev. Lett.}, 99(4):048102, July 2007.

\bibitem{Thutupalli2011}
S.~Thutupalli, R.~Seemann, and S.~Herminghaus.
\newblock Swarming behavior of simple model squirmers.
\newblock {\em N. J. Phys.}, 13(7):073021, July 2011.

\bibitem{Izri2014}
Z.~Izri, M.~N. van~der Linden, S.~Michelin, and O.~Dauchot.
\newblock Self-propulsion of pure water droplets by spontaneous
  {M}arangoni-stress-driven motion.
\newblock {\em Phys. Rev. Lett.}, 113(24):248302, December 2014.

\bibitem{Moerman2017}
P.~G. Moerman, H.~W. Moyses, E.~B. van~der Wee, D.~G. Grier, A.~van Blaaderen,
  W.~K. Kegel, J.~Groenewold, and J.~Brujic.
\newblock Solute-mediated interactions between active droplets.
\newblock {\em Phys. Rev. E}, 96(3):032607, September 2017.

\bibitem{Golestanian2007}
R.~Golestanian, T.~B. Liverpool, and A.~Ajdari.
\newblock Designing phoretic micro- and nano-swimmers.
\newblock {\em N. J. Phys.}, 9(5):126--126, May 2007.

\bibitem{Maass2016}
C.~C. Maass, C.~Kr\"{u}ger, S.~Herminghaus, and C.~Bahr.
\newblock Swimming droplets.
\newblock {\em Ann. Rev. Cond. Matt. Phys.}, 7(1):171--193, March 2016.

\bibitem{Moran2017}
J.~L. Moran and J.~D. Posner.
\newblock Phoretic self-propulsion.
\newblock {\em Ann. Rev. Fluid Mech.}, 49(1):511--540, January 2017.

\bibitem{Paxton2004}
W.~F. Paxton, K.~C. Kistler, C.~C. Olmeda, A.~Sen, S.~K.~St. Angelo, Y.~Cao,
  T.~E. Mallouk, P.~E. Lammert, and V.~H. Crespi.
\newblock Catalytic nanomotors:~ autonomous movement of striped nanorods.
\newblock {\em J. Am. Chem. Soc.}, 126(41):13424--13431, October 2004.

\bibitem{Kummel2013}
F.~K\"{u}mmel, B.~ten Hagen, R.~Wittkowski, I.~Buttinoni, R.~Eichhorn,
  G.~Volpe, H.~L\"{o}wen, and C.~Bechinger.
\newblock Circular motion of asymmetric self-propelling particles.
\newblock {\em Phys. Rev. Lett.}, 110(19):198302, May 2013.

\bibitem{Michelin2015}
S.~Michelin and E.~Lauga.
\newblock Autophoretic locomotion from geometric asymmetry.
\newblock {\em Eur. Phys. J. E}, 38(2), February 2015.

\bibitem{Varma2018}
A.~Varma, T.~D. Montenegro-Johnson, and S.~Michelin.
\newblock Clustering-induced self-propulsion of isotropic autophoretic
  particles.
\newblock {\em Soft Matter}, 14(35):7155--7173, 2018.

\bibitem{Yu2018}
T.~Yu, P.~Chuphal, S.~Thakur, S.~Y. Reigh, D.~P. Singh, and P.~Fischer.
\newblock Chemical micromotors self-assemble and self-propel by spontaneous
  symmetry breaking.
\newblock {\em Chem. Comm.}, 54(84):11933--11936, 2018.

\bibitem{Morozov2020}
M.~Morozov.
\newblock Adsorption inhibition by swollen micelles may cause multistability in
  active droplets.
\newblock {\em Soft Matter}, 16(24):5624--5632, 2020.

\bibitem{Hokmabad2021}
B.~V. Hokmabad, R.~Dey, M.~Jalaal, D.~Mohanty, M.~Almukambetova, K.~A. Baldwin,
  D.~Lohse, and C.~C. Maass.
\newblock Emergence of bimodal motility in active droplets.
\newblock {\em Phys. Rev. X}, 11(1):011043, March 2021.

\bibitem{Morozov2019a}
M.~Morozov and S.~Michelin.
\newblock Self-propulsion near the onset of {M}arangoni instability of
  deformable active droplets.
\newblock {\em J. Fluid Mech.}, 860:711--738, January 2019.

\bibitem{Suga2018}
M.~Suga, S.~Suda, M.~Ichikawa, and Kimura Y.
\newblock Self-propelled motion switching in nematic liquid crystal droplets in
  aqueous surfactant solutions.
\newblock {\em Phys. Rev. E}, 97(6):062703, June 2018.

\bibitem{Hu2019}
W.-F. Hu, T.-S. Lin, S.~Rafai, and C.~Misbah.
\newblock Chaotic swimming of phoretic particles.
\newblock {\em Phys. Rev. Lett.}, 123(23):238004, December 2019.

\bibitem{Morozov2019b}
M.~Morozov and S.~Michelin.
\newblock Nonlinear dynamics of a chemically-active drop: From steady to
  chaotic self-propulsion.
\newblock {\em J. Chem. Phys.}, 150(4):044110, January 2019.

\bibitem{Michelin2013}
S.~Michelin, E.~Lauga, and D.~Bartolo.
\newblock Spontaneous autophoretic motion of isotropic particles.
\newblock {\em Phys. Fluids}, 25(6):061701, June 2013.

\bibitem{deBlois2021}
C.~de~Blois, V.~Bertin, S.~Suda, M.~Ichikawa, M.~Reyssat, and O.~Dauchot.
\newblock {Swimming droplet in 1D geometries, an active Bretherton problem}.
\newblock {\em Soft Matter}, 17:6646--6660, 2021.

\bibitem{Cheon2021}
S.~I. Cheon, L.~B.~Capaverde Silva, A.~S. Khair, and L.~D. Zarzar.
\newblock Interfacially-adsorbed particles enhance the self-propulsion of oil
  droplets in aqueous surfactant.
\newblock {\em Soft Matter}, 17(28):6742--6750, 2021.

\bibitem{Kruger2016b}
C.~Kr\"{u}ger, C.~Bahr, S.~Herminghaus, and C.~C. Maass.
\newblock Dimensionality matters in the collective behaviour of active
  emulsions.
\newblock {\em Eur. Phys. J. E}, 39(6), June 2016.

\bibitem{deBlois2019}
C.~de~Blois, M.~Reyssat, S.~Michelin, and O.~Dauchot.
\newblock Flow field around a confined active droplet.
\newblock {\em Phys. Rev. Fluids}, 4(5):054001, May 2019.

\bibitem{Lippera2020}
K.~Lippera, M.~Morozov, M.~Benzaquen, and S.~Michelin.
\newblock Collisions and rebounds of chemically active droplets.
\newblock {\em J. Fluid Mech.}, 886, January 2020.

\bibitem{Illien2020}
P.~Illien, C.~de~Blois, Y.~Liu, M.~N. van~der Linden, and O.~Dauchot.
\newblock Speed-dispersion-induced alignment: A one-dimensional model inspired
  by swimming droplets experiments.
\newblock {\em Phys. Rev. E}, 101(4):040602, April 2020.

\bibitem{Kim1991}
S.~Kim and S.~J. Karrila.
\newblock {\em Microhydrodynamics: Principles and Selected Applications}.
\newblock Butterworth-Heineman, Boston, 1991.

\bibitem{Zhu2013}
L.~Zhu, E.~Lauga, and L.~Brandt.
\newblock Low-{R}eynolds-number swimming in a~capillary~tube.
\newblock {\em J. Fluid Mech.}, 726:285--311, May 2013.

\bibitem{MontenegroJohnson2015}
T.~D. Montenegro-Johnson, S.~Michelin, and E.~Lauga.
\newblock A regularised singularity approach to phoretic problems.
\newblock {\em Eur. Phys. J. E}, 38(12), December 2015.

\bibitem{Yan2016}
W.~Yan and J.~F. Brady.
\newblock The behavior of active diffusiophoretic suspensions: An accelerated
  {L}aplacian dynamics study.
\newblock {\em J. Chem. Phys.}, 145(13):134902, October 2016.

\bibitem{RojasPerez2021}
F.~Rojas-P{\'{e}}rez, B.~Delmotte, and S.~Michelin.
\newblock Hydrochemical interactions of phoretic particles: a regularized
  multipole framework.
\newblock {\em J. Fluid Mech.}, 919, May 2021.

\bibitem{Lippera2020b}
K.~Lippera, M.~Benzaquen, and S.~Michelin.
\newblock Bouncing, chasing, or pausing: Asymmetric collisions of active
  droplets.
\newblock {\em Phys. Rev. Fluids}, 5(3):032201, March 2020.

\bibitem{Johansen1998}
H.~Johansen and P.~Colella.
\newblock A {C}artesian grid embedded boundary method for
  {P}oisson{\textquotesingle}s equation on irregular domains.
\newblock {\em J. Comp. Phys.}, 147(1):60--85, November 1998.

\bibitem{Schwartz2006}
P.~Schwartz, M.~Barad, P.~Colella, and T.~Ligocki.
\newblock A {C}artesian grid embedded boundary method for the heat equation and
  {P}oisson's equation in three dimensions.
\newblock {\em J. Comp. Phys.}, 211(2):531--550, January 2006.

\bibitem{Popinet2015}
S.~Popinet.
\newblock A quadtree-adaptive multigrid solver for the
  {S}erre{\textendash}{G}reen{\textendash}{N}aghdi equations.
\newblock {\em J. Comp. Phys.}, 302:336--358, December 2015.

\bibitem{Sherwood2018}
J.~D. Sherwood and S.~Ghosal.
\newblock Nonlinear electrophoresis of a tightly fitting sphere in a
  cylindrical tube.
\newblock 843:847--871, March 2018.

\bibitem{Selcuk2020}
C.~SelÃ§uk, A.~R. Ghigo, S.~Popinet, and A.~Wachs.
\newblock A fictitious domain method with distributed {L}agrange multipliers on
  adaptive quad/octrees for the direct numerical simulation of particle-laden
  flows.
\newblock {\em J. Comp. Phys.}, 430:109954, 2021.

\bibitem{Sangani1996}
A.~S. Sangani and G.~Mo.
\newblock An {O}({N}) algorithm for {S}tokes and {L}aplace interactions of
  particles.
\newblock {\em Phys. Fluids}, 8(8):1990--2010, August 1996.

\bibitem{Pozrikidis1992}
C.~Pozrikidis.
\newblock {\em Boundary Integral and Singularity Methods for Linearized Viscous
  Flow}.
\newblock Cambridge University Press, February 1992.

\bibitem{Delmotte2015}
B.~Delmotte, E.~E. Keaveny, F.~Plourabou{\'{e}}, and E.~Climent.
\newblock Large-scale simulation of steady and time-dependent active
  suspensions with the force-coupling method.
\newblock {\em J. Comp. Phys.}, 302:524--547, December 2015.

\bibitem{Popinet2003}
S.~Popinet.
\newblock {G}erris: a tree-based adaptive solver for the incompressible {E}uler
  equations in complex geometries.
\newblock {\em J. Comp. Phys.}, 190(2):572--600, September 2003.

\bibitem{Bell1989}
J.~B. Bell, P.~Colella, and H.~M. Glaz.
\newblock A second-order projection method for the incompressible
  {N}avier-{S}tokes equations.
\newblock {\em J. Comp. Phys.}, 85(2):257--283, December 1989.

\bibitem{Schneiders2016}
L~Schneiders, C.~G\"{u}nther, M.~Meinke, and W.~Schr\"{o}der.
\newblock An efficient conservative cut-cell method for rigid bodies
  interacting with viscous compressible flows.
\newblock {\em J. Comp. Phys.}, 311:62--86, April 2016.

\bibitem{vanHooft2018}
J.~Antoon van Hooft, St{\'{e}}phane Popinet, Chiel~C. van Heerwaarden, Steven
  J.~A. van~der Linden, Stephan~R. de~Roode, and Bas J.~H. van~de Wiel.
\newblock Towards adaptive grids for atmospheric boundary-layer simulations.
\newblock {\em Boundary-Layer Meteorol.}, 167(3):421--443, February 2018.

\bibitem{Stone1996}
H.~A. Stone and A.~D.~T. Samuel.
\newblock Propulsion of microorganisms by surface distortions.
\newblock {\em Phys. Rev. Lett.}, 77(19):4102--4104, November 1996.

\bibitem{Anderson1982}
J.~L. Anderson, M.~E. Lowell, and D.~C. Prieve.
\newblock Motion of a particle generated by chemical gradients part 1.
  non-electrolytes.
\newblock {\em Journal of Fluid Mechanics}, 117:107--121, April 1982.

\bibitem{Desai2021}
N.~Desa\"i and S.~Michelin.
\newblock Instability and self-propulsion of active droplets along a wall.
\newblock under review, 2021.

\bibitem{Leal2007}
L.~Gary Leal.
\newblock {\em Advanced Transport Phenomena}.
\newblock Cambridge University Press, 2007.

\bibitem{Jin2019}
Chenyu Jin, J{\'{e}}r{\'{e}}my Vachier, Soumya Bandyopadhyay, Tamara
  Macharashvili, and Corinna~C. Maass.
\newblock Fine balance of chemotactic and hydrodynamic torques: When
  microswimmers orbit a pillar just once.
\newblock {\em Physical Review E}, 100(4), October 2019.

\bibitem{Hokmabad2020b}
Babak~Vajdi Hokmabad, Suropriya Saha, Jaime Agudo-Canalejo, Ramin Golestanian,
  and Corinna~C. Maass.
\newblock Quantitative characterization of chemorepulsive alignment-induced
  interactions in active emulsions, 2020.

\end{thebibliography}
\providecommand{\noopsort}[1]{}\providecommand{\singleletter}[1]{#1}%

\end{document}